\documentclass[english]{paper}
\usepackage[T1]{fontenc}
\usepackage[latin9]{inputenc}
\usepackage{geometry}
\usepackage{color}
\usepackage{babel}
\usepackage{array}
\usepackage{float}
\usepackage{mathrsfs}
\usepackage{multirow}
\usepackage{amsmath}
\usepackage{amssymb}
\usepackage[unicode=true,pdfusetitle,
 bookmarks=true,bookmarksnumbered=false,bookmarksopen=false,
 breaklinks=true,pdfborder={0 0 0},pdfborderstyle={},backref=false,colorlinks=true]
 {hyperref}
\usepackage{breakurl}

\makeatletter

\providecommand{\tabularnewline}{\\}

\numberwithin{figure}{section}
\numberwithin{equation}{section}
\newcommand{\lyxaddress}[1]{
\par {\raggedright #1
\vspace{1.4em}
\noindent\par}
}

\@ifundefined{date}{}{\date{}}
\makeatother

\begin{document}

\title{{\Large{}Quaternionic approach to dual Magneto-hydrodynamics of dyonic
cold plasma}}

\author{\textrm{\textbf{\textup{\large{}B. C. Chanyal}}}\thanks{\textit{Corresponding Author (email: bcchanyal@gmail.com, bcchanyal@gbpuat.ac.in)}}\textrm{\textbf{\textup{\large{}
~and Mayank Pathak}}}}
\maketitle

\lyxaddress{\textit{Department of Physics, G. B. Pant University of Agriculture
\& Technology, Pantnagar-263145 (Uttarakhand), India}}
\begin{abstract}
The dual magneto-hydrodynamics of dyonic plasma describes the study
of electrodynamics equations along with the transport equations in
the presence of electrons and magnetic monopoles. In this paper, we
formulate the quaternionic dual fields equations, namely, the hydro-electric
and hydro-magnetic fields equations which are an analogous to the
generalized Lamb vector field and vorticity field equations of dyonic
cold plasma fluid. Further, we derive the quaternionic Dirac-Maxwell
equations for dual magneto-hydrodynamics of dyonic cold plasma. We
also obtain the quaternionic dual continuity equations that describe
the transport of dyonic fluid. Finally, we establish an analogy of
Alfven wave equation which may generate from the flow of magnetic
monopoles in the dyonic field of cold plasma. The present quaternionic
formulation for dyonic cold plasma is well invariant under the duality,
Lorentz and CPT transformations.

\end{abstract}

\section{Introduction }

In the past few decades, astronomers predicted that the universe was
composed almost entirely of the baryonic matter (ordinary matter).
According to Bachynski \cite{key-1}, more than 99\% of the matter
in the universe is in plasma state. This type of matter may be consist
of baryonic and non-baryonic matter. The first experimental evidence
of the existence of plasma was given by American Physicists \cite{key-2}.
In plasma, consisting of charged and neutral particles, the inter-ionic
force between particles show electromagnetic in nature. Therefore,
due to the long range order of Coulomb force charged particles interact
with all other charged particles resulting in a collective behavior
of plasma. In 1942, Alfven \cite{key-3} gave the theory of Magnetohydrodynamics
(MHD) and suggested that electrically conducting fluid can support
the propagation of shear waves called the \textit{Alfven waves}. Basically,
MHD describes the behavior of electrically conducting fluid in the
presence of magnetic field \cite{key-4}. It is macroscopic theory
that assumes the electrons, ions and charged particles moves together
and treated them as a single fluid component known as single-fluid
theory. The plasma along with MHD is simply described by a single
temperature, velocity and density. However, when the MHD wave propagates
faster than plasma thermal speed then the effect of temperature can
be neglected \cite{key-5}. This is called a cold plasma approximation
(i.e., in cold plasma approximation, temperature doesn\textquoteright t
take into account). In this approximation, there is no wave related
to pressure fluctuation (e.g. sound waves). On the other hand, the
hot and warm plasmas are another sates of plasma where the collision
between electrons and gas molecules are so frequent that there is
a thermal equilibrium between electron and the gas molecules. 

Meyer-Vernet \cite{key-6} discussed the role of magnetic monopole
in conducting fluid (plasma). The magnetic monopole proposed by Dirac
\cite{key-7}, it is a hypothetical elementary particle having only
one magnetic pole. Dirac also pointed out that if there exist any
monopole in the universe then all the electric charge in the universe
will be quantized \cite{key-8}. Schwinger \cite{key-9,key-10}, an
exception to the argument against the existence of monopole, and formulated
relativistically covariant quantum field theory of magnetic monopoles
which maintained complete symmetry between electric and magnetic fields.
Therefore, the name of particles that carrying simultaneously the
electric and magnetic charges called Dyons. Further, the theoretical
approach of Schwinger \cite{key-9,key-10} and Zwanziger \cite{key-11}
describe the theory of dyonic particles. Peres \cite{key-12} pointed
out the controversial nature \cite{key-13} of the singular lines
of magnetic monopoles and established the charged quantization condition
in purely group theoretical manner without using them. In view of
mathematical physics, the study of four dimensional particles (dyons)
in distinguish mediums can be explain by division algebras. There
are four types of divisions algebras \cite{key-14}, namely the real,
complex, quaternion and octonion algebras. The complex algebra is
an extension of real numbers, the quaternion is an extension of complex
numbers while the octonion is an extension of quaternions. Quaternionic
algebra \cite{key-15} can also express by the four-dimensional Euclidean
spaces \cite{key-16,key-17}, and it has vast applications in the
multiple branches of physics.

Further, Rajput \cite{key-18} pointed out an effective unified theory
for quaternionic generalized electromagnetic and gravitational fields
of dyons by using the quaternion algebra. The quaternionic form of
classical and quantum electrodynamics have been already discussed
\cite{key-19,key-20,key-21,key-22}. Many authors \cite{key-23,key-24,key-25,key-26,key-27,key-28,key-29}
have studied the role of hyper-complex algebras in various branches
of physics. Recently, Chanyal \cite{key-30,key-31} independently
proposed a novel approach on the quaternionic covariant theory for
relativistic quantum mechanics, and established the quantized Dirac-Maxwell
equations for dyons. Besides, in literature \cite{key-32,key-33,key-34},
the reformulation of incompressible plasma fluids and MHD equations
have been discussed in terms of hyper-complex numbers. Keeping in
view the importance of quaternionic algebras, we establish the MHD
field equations for dyonic cold plasma. Starting with the definitions
of one-fluid and two-fluid theory of plasma, we identify the cold
plasma approximation where the thermal effects (or pressure effects)
of conducting fluid will be neglected. Further, we introduce the dual
MHD equations of dyonic plasma consisted with electrons, magnetic
monopoles and their counter partners viz. ions and magneto-ions. In
this study, we clarify that the dominating aspect for the dyonic cold
plasma approximation is the dynamics of electrons along with magnetic
monopoles. As we know that the generalized Dirac-Maxwell like equations
are primary equations to explain the dynamics of dyonic cold plasma.
Therefore, undertaking the quaternionic dual-velocity and dual-enthalpy
of dyonic cold plasma, we have made an attempt to formulate the quaternionic
hydro-electric and hydro-magnetic fields equations, which are an analogous
to the generalized Lamb vector field and vorticity field of conducting
dyonic fluid. The Lorenz gauge conditions for dyonic cold plasma fluid
are also obtained. Further, we derive the generalized quaternionic
Dirac-Maxwell equations to the case of dual magneto-hydrodynamics
of dyonic cold plasma. We have discussed that these Dirac-Maxwell
equations for dyonic cold plasma are well invariant under the duality,
Lorentz and CPT transformations. Finally, the Alfven wave like equation
is established which may propagate from the flow of magnetic monopoles
in the dyonic cold plasma.

\section{The quaternions}

Through the extension of the set of natural numbers to the integers,
a complex number $\mathbb{C}$ is defined by the set of all real linear
combinations of the unit elements $(1,\,i)$, such that
\begin{align}
\mathbb{C}\,\longmapsto\, & \left\{ \alpha=\alpha_{1}+i\alpha_{2}\,\mid\,(\alpha_{1},\,\alpha_{2}\in\mathbb{R})\right\} \,,\label{eq:1}
\end{align}
where the real number $\alpha_{1}$ is called the real part and $\alpha_{2}$
is called the imaginary part of a complex number$\alpha$. If the
real part $Re(\alpha)=0,$ then we can say that $\alpha$ is purely
imaginary. As such, the Euclidean scalar product as $\mathbb{C}\times\mathbb{C}\longmapsto\mathbb{R}$
is then defined by
\begin{align}
\left\langle \alpha,\,\beta\right\rangle \,= & \,\,Re\,(\alpha\cdot\bar{\beta})=\alpha_{1}\beta_{1}+\alpha_{2}\beta_{2}\,,\label{eq:2}
\end{align}
where $\alpha=\alpha_{1}+i\alpha_{2}$ and $\beta=\beta_{1}+i\beta_{2}$
are two complex numbers. The modulus of any complex number is also
defined by $\mid\alpha\mid=\sqrt{\alpha\cdot\bar{\alpha}}=\sqrt{\alpha_{1}^{2}+\alpha_{2}^{2}}$.

However, a complex field $\mathbb{C}$ is a finite dimensional real
vector space, so that we can easily extend the complex number into
the quaternionic field $\mathbb{H}$ by losing the commutativity of
multiplication. Thus, the quaternion represents the natural extension
of complex numbers and form an algebra under addition and multiplication.
Hamilton \cite{key-15} described a four-dimensional quaternionic
algebra and applied it to mechanics in three-dimensional space. A
striking feature of quaternions is that the product of two quaternions
is non-commutative, meaning that the product of two quaternions depends
on which factor is to the left of the multiplication sign and which
factor is to the right. 

Thus the allowed four dimensional Hamilton vector space is defined
by quaternion algebra $\mathbb{H}$ over the field of real numbers
$\mathbb{R}$ as
\begin{align}
\mathbb{H}\,\,\longmapsto & \,\,\left\{ \alpha=\,\,\sum_{j=0}^{3}e_{j}\alpha_{j}=\,\,e_{0}\alpha_{0}+e_{1}\alpha_{1}+e_{2}\alpha_{2}+e_{3}\alpha_{3}\,\mid\,\forall\,\alpha_{j}\in\mathbb{R}\right\} \,,\label{eq:3}
\end{align}
where the Hamilton vector space ($\mathbb{H}$) has the quaternionic
elements ($e_{0}$, $e_{1}$, $e_{2}$, $e_{3}$), are called quaternion
basis elements while $\alpha_{0}$, $\alpha_{1}$, $\alpha_{2}$,
$\alpha_{3}$ are the real quarterate of a quaternion. As such the
addition of two quaternions $\alpha=e_{0}\alpha_{0}+e_{1}\alpha_{1}+e_{2}\alpha_{2}+e_{3}\alpha_{3}$
and $\beta=e_{0}\beta_{0}+e_{1}\beta_{1}+e_{2}\beta_{2}+e_{3}\beta_{3}$
is given by
\begin{align}
\alpha+\beta\,= & \,\,e_{0}\left(\alpha_{0}+\beta_{0}\right)+e_{1}\left(\alpha_{1}+\beta_{1}\right)+e_{2}\left(\alpha_{2}+\beta_{2}\right)+e_{3}\left(\alpha_{3}+\beta_{3}\right)\,\,,\,\,\,\,\,\,\forall\,\left(\alpha,\,\beta\right)\in\mathbb{H}\,.\label{eq:4}
\end{align}
Here, the quaternionic addition is clearly associative and commutative.
The additive identity element is defined by the zero element, i.e.,
\begin{align}
0\,\,= & \,\,e_{0}0+e_{1}0+e_{2}0+e_{3}0\,,\label{eq:5}
\end{align}
and the additive inverse of $\alpha\in\mathbb{H}$ is given by
\begin{align}
-\alpha\,\,= & \,\,e_{0}\left(-\alpha_{0}\right)+e_{1}\left(-\alpha_{1}\right)+e_{2}\left(-\alpha_{2}\right)+e_{3}\left(-\alpha_{3}\right)\,.\label{eq:6}
\end{align}
Correspondingly, the product of two quaternions, i.e. $\left(\alpha\circ\beta\right)\in\mathbb{H}$
can be expressed by
\begin{align}
\alpha\circ\beta\,\,= & \,\,e_{0}\left(\alpha_{0}\beta_{0}-\alpha_{1}\beta_{1}-\alpha_{2}\beta_{2}-\alpha_{3}\beta_{3}\right)\nonumber \\
+ & e_{1}\left(\alpha_{0}\beta_{1}+\alpha_{1}\beta_{0}+\alpha_{2}\beta_{3}-\alpha_{3}\beta_{2}\right)\nonumber \\
+ & e_{2}\left(\alpha_{0}\beta_{2}-\alpha_{1}\beta_{3}+\alpha_{2}\beta_{0}+\alpha_{3}\beta_{1}\right)\nonumber \\
+ & e_{3}\left(\alpha_{0}\beta_{3}+\alpha_{1}\beta_{2}-\alpha_{2}\beta_{1}+\alpha_{3}\beta_{0}\right)\,.\label{eq:7}
\end{align}
We may notice that this quaternionic product is associative, but not
commutative. The quaternionic unit elements $(e_{0}\,,e_{1},\,e_{2},\,e_{3})$
are followed the given relations,
\begin{align}
e_{0}^{2} & =\,1\,,\,\,e_{A}^{2}=\,-1\,,\nonumber \\
e_{0}e_{A} & =\,e_{A}e_{0}=e_{A}\,,\nonumber \\
e_{A}e_{B} & =-\delta_{AB}e_{0}+f_{ABC}e_{C}\,,\,\,\,(\forall\,A,B,C=1,2,3)\,\label{eq:8}
\end{align}
where $\delta_{AB}$ is the delta symbol and $f_{ABC}$ is the Levi
Civita three-index symbol having value $f_{ABC}=+1$ for cyclic permutation,
$f_{ABC}=-1$ for anti-cyclic permutation and $f_{ABC}=0$ for any
two repeated indices. Further, we also may write the following relations
to quaternion basis elements 
\begin{align}
\left[e_{A},\,\,e_{B}\right] & \,=\,2\,f_{ABC}\,e_{C}\,,\nonumber \\
\left\{ e_{A},\,\,e_{B}\right\}  & \,=\,-2\,\delta_{AB}e_{0}\,,\nonumber \\
e_{A}(\,e_{B}\,e_{C}) & \,=\,(e_{A}\,e_{B}\,)\,e_{C}\,,\label{eq:9}
\end{align}
where the brackets $[\,\,,\,\,]$ and $\{\,\,,\,\,\}$ are used respectively
for commutation and the anti-commutation relations. Thus the above
multiplication rules governed the ordinary dot and cross product,
i.e.,
\begin{align}
\alpha\circ\beta\,\,= & \,\,\left(\alpha_{0}\beta_{0}-\boldsymbol{\alpha}\cdot\boldsymbol{\beta},\,\,\,\alpha_{0}\boldsymbol{\beta}+\beta_{0}\boldsymbol{\alpha}+(\boldsymbol{\alpha}\times\boldsymbol{\beta})\right)\,,\label{eq:10}
\end{align}
where we take $\boldsymbol{\alpha}\times\boldsymbol{\beta}\neq0$
for non-commutative product of quaternion. The quaternionic product
with the scalar quantity $\xi$ is given by
\begin{align}
\xi\circ\alpha\,\,= & \,\,e_{0}\left(\xi\alpha_{0}\right)+e_{1}\left(\xi\alpha_{1}\right)+e_{2}\left(\xi\alpha_{2}\right)+e_{3}\left(\xi\alpha_{3}\right)\,.\label{eq:11}
\end{align}
 As such, the multiplication identity element can expressed by the
unit elements,
\begin{align}
1\,\,= & \,\,e_{0}1+e_{1}0+e_{2}0+e_{3}0\,.\label{eq:12}
\end{align}
Moreover, a quaternion can also be decomposed in terms of scalar $(S(\alpha))$
and vector $(\boldsymbol{V}(\alpha))$ parts as 
\begin{align}
S(\alpha)\, & =\,\frac{1}{2}(\,\alpha\,+\,\bar{\alpha}\,)\,,\label{eq:13}\\
\boldsymbol{V}(\alpha)\, & =\,\frac{1}{2}(\,\alpha\,-\,\bar{\alpha}\,)\,,\label{eq:14}
\end{align}
where the quaternionic conjugate $\bar{\alpha}$ is expressed by
\begin{align}
\bar{\alpha}\,\,= & \,\,e_{0}\alpha_{0}-\left(e_{1}\alpha_{1}+e_{2}\alpha_{2}+e_{3}\alpha_{3}\right)\,.\label{eq:15}
\end{align}
The real and imaginary parts of $\alpha$ can be written as
\begin{align}
Re(\mathbb{H})\,\,= & \,\,\alpha_{0}\,,\label{eq:16}\\
Im(\mathbb{H})\,\,= & \,\,\left\{ e_{1}\alpha_{1}+e_{2}\alpha_{2}+e_{3}\alpha_{3}\,\mid\,\forall\,\alpha_{j=1,2,3}\in\mathbb{R}\right\} \,\subseteq\mathbb{H}\,.\label{eq:17}
\end{align}
If $Re(\mathbb{H})=0$ and $\alpha\neq0,$ then $\alpha$ is said
to be purely imaginary quaternions. Therefore, all quaternions with
zero real is simplified as imaginary space of $\mathbb{H}$, where
the imaginary space $Im(\mathbb{H})\in\mathbb{R}^{3}$ is a three
dimensional real vector space,
\begin{align}
Im(\alpha)\,\,=\, & \left(\alpha_{1},\,\alpha_{2},\,\alpha_{3}\right)\,\Longrightarrow\,Im(\alpha)^{\dagger}\,=\,\,\begin{pmatrix}\alpha_{1}\\
\alpha_{2}\\
\alpha_{3}
\end{pmatrix}\,.\label{eq:18}
\end{align}
Interestingly, we may write the following form of quaternion as
\begin{align}
\alpha\,= & \,\,Re(\alpha)+\sum_{j=1}^{3}e_{j}\,Im(\alpha_{j})\,.\label{eq:19}
\end{align}
The quaternionic Euclidean scalar product $\mathbb{H}\times\mathbb{H}\longmapsto\mathbb{R}$
can also be expressed as
\begin{align}
\left\langle \alpha,\,\beta\right\rangle \,=\,\,Re\,(\alpha\circ\bar{\beta}) & \,=\,\alpha_{0}\beta_{0}+\alpha_{1}\beta_{1}+\alpha_{2}\beta_{2}+\alpha_{3}\beta_{3}\,.\label{eq:20}
\end{align}
Like complex numbers, the modulus of quaternion $\alpha$ is then
defined as
\begin{align}
\mid\alpha\mid\,= & \,\sqrt{\alpha_{0}^{2}+\alpha_{1}^{2}+\alpha_{2}^{2}+\alpha_{3}^{2}}\,.\label{eq:21}
\end{align}
Since, there exists the norm $N(\alpha)=\alpha\circ\bar{\alpha}$
of a quaternion, we have a division i.e., every $\alpha$ has an inverse
of a quaternion and is expressed as
\begin{align}
\alpha^{-1}\,= & \,\frac{\bar{\alpha}}{\mid\alpha\mid}\,.\label{eq:22}
\end{align}
While the quaternion conjugation satisfies the following property
\begin{align}
\overline{\alpha_{1}\circ\alpha_{2}}\,\,= & \,\,\overline{\alpha_{1}}\,\circ\,\overline{\alpha_{2}}.\label{eq:23}
\end{align}
The norm of the quaternion is positive definite and obey the composition
law
\begin{align}
N\left(\alpha_{1}\circ\alpha_{2}\right)\,= & \,N\left(\alpha_{1}\right)\,\circ\,N\left(\alpha_{2}\right).\label{eq:24}
\end{align}
The quaternion elements are non-Abelian in nature and thus represent
a non-commutative division ring. Quaternion is an important fundamental
mathematical tool that appropriate for four-dimensional world.

\section{Magneto-hydrodynamics of cold plasma}

Let us start with the basic parameters of the plasma. As we know that
the plasma exists in many more forms in nature which has a wide spread
use in the science and technology. The theory of plasma is divided
into three categories \cite{key-35}, namely, the microscopic theory,
kinetic theory and the fluid theory. In briefly, the microscopic theory
is based on the motion of all the individual particles (e.g. electrons,
ions, atoms, molecules, radicals, etc). According to Klimontovich
\cite{key-36}, the time evolution of the particle density ($\rho_{s}\longmapsto\rho_{s}(\boldsymbol{r},\boldsymbol{v},t)$)
is expressed by
\begin{align}
\frac{\partial\rho_{s}}{\partial t}+\boldsymbol{v}\cdot\boldsymbol{\nabla}\rho_{s}+\frac{q_{s}}{m_{s}}(\boldsymbol{E}+\boldsymbol{v}\times\boldsymbol{B})\cdot\boldsymbol{\nabla}\rho_{s} & \,=\,\,0\,,\label{eq:25}
\end{align}
where \textbf{$\boldsymbol{v}$} is the velocity of particles, ($q_{s}$,
$m_{s}$) are the effective charge and mass of the $s-$species particles
and ($\boldsymbol{E}$, $\boldsymbol{B}$) are the electric and magnetic
field produced by the microscopic particles. Besides, the collisionless
kinetic theory of plasma proposed by Vlasov \cite{key-37},\textbf{
}which has included the Boltzmann distribution function $f_{s}\,\simeq\,\,\left\langle \rho_{s}\right\rangle $
as \cite{key-35},
\begin{align}
\frac{\partial f_{s}}{\partial t}+\boldsymbol{v}.\boldsymbol{\nabla}f_{s}+\frac{q_{s}}{m_{s}}(\boldsymbol{E}+\boldsymbol{v}\times\boldsymbol{B}).\boldsymbol{\nabla}f_{s} & \,=\,\,0\,\,.\label{eq:26}
\end{align}
In equations (\ref{eq:25}) and (\ref{eq:26}), we may consider that
the two dominating particles (i.e. electrons and ions both) constitute
the dynamics of plasma, called the two-fluid theory of plasma \cite{key-35,key-36,key-37,key-38}.
For the two-fluid theory of plasma, at a given position ($x$) the
mass and charge densities become
\begin{align}
\rho_{M}(x)\,\,= & \,\,\,m_{e}n_{e}(x)+m_{i}n_{i}(x)\,,\label{eq:27}\\
\rho_{c}(x)\,\,= & \,\,\,q_{e}n_{e}(x)+q_{i}n_{i}(x)\,,\label{eq:28}
\end{align}
where $m_{e},\,n_{e},\,\text{and}\,q_{e}$ are defined the mass, total
number and charge of electrons while $m_{i},\,n_{i},\,\text{and}\,q_{i}$
are defined the mass, total number and charge of ions, respectively.
The center of mass fluid velocity can be expressed as
\begin{align}
\boldsymbol{v}\,\,= & \,\,\,\frac{1}{\rho_{M}(x)}\left(\boldsymbol{v}_{e}m_{e}n_{e}(x)+\boldsymbol{v}_{i}m_{i}n_{i}(x)\right)\,,\label{eq:29}
\end{align}
and the current density becomes
\begin{align}
\boldsymbol{J}\,\,= & \,\,\,q_{e}n_{e}\boldsymbol{v}_{e}+q_{i}n_{i}\boldsymbol{v}_{i}\,\,.\label{eq:30}
\end{align}
The continuity equations can be written as

\begin{align}
\frac{\partial\rho_{M}}{\partial t}+\boldsymbol{\nabla}\cdot(\rho_{M}\boldsymbol{v}) & \,=\,\,0\,,\,\,\,\,\,\,\,(\text{mass conservation law})\label{eq:31}\\
\frac{\partial\rho_{c}}{\partial t}+\boldsymbol{\nabla}\cdot\boldsymbol{J} & \,=\,\,0\,,\,\,\,\,\,\,\,(\text{charge conservation law})\label{eq:32}
\end{align}
As such, the momentum equation for plasma fluid is expressed as \cite{key-35},
\begin{align}
\rho_{M}\left(\frac{\partial}{\partial t}+\boldsymbol{v}\cdot\boldsymbol{\nabla}\right)\boldsymbol{v} & \,\,=\,\,\left(\boldsymbol{J\times B}\right)+\rho_{c}\boldsymbol{E}-\boldsymbol{\nabla}p\,,\label{eq:33}
\end{align}
where $\boldsymbol{\nabla}p$ is the pressure force introduced due
to the inhomogeneity of the plasma and $\left(\boldsymbol{J}\times\boldsymbol{B}\right)$
is a Lorentz force per unit volume element. Now, we introduce an acceleration
to the conducting fluid,
\begin{align}
\frac{\partial\boldsymbol{v}}{\partial t}\,\,\longmapsto\,\, & \left(\frac{\partial}{\partial t}+\boldsymbol{v}\cdot\boldsymbol{\nabla}\right)\boldsymbol{v}\,,\label{eq:34}
\end{align}
where the term $\left(\boldsymbol{v}\cdot\boldsymbol{\nabla}\right)\boldsymbol{v}$
is used for the convective acceleration of fluid. Furthermore, the
generalized Ohm's law becomes \cite{key-35}
\begin{align}
\frac{m_{e}m_{i}}{\rho_{M}\,e^{2}}\frac{\partial\boldsymbol{J}}{\partial t} & \,\,=\,\,\frac{m_{i}}{2\rho_{M}\,e}\boldsymbol{\nabla}p+\boldsymbol{E}+\left(\boldsymbol{v}\times\boldsymbol{B}\right)-\frac{m_{i}}{\rho_{M}\,e}\left(\boldsymbol{J}\times\boldsymbol{B}\right)-\frac{\boldsymbol{J}}{\sigma}\,,\label{eq:35}
\end{align}
where $\sigma$ denotes the conductivity of fluid. One can define
the Maxwell's equations with natural unit ($\hbar=c=1$) as,
\begin{align}
\boldsymbol{\nabla\cdot E}\, & =\,\rho_{c}\,,\label{eq:36}\\
\boldsymbol{\nabla\cdot B}\, & =\,0\,,\label{eq:37}\\
\boldsymbol{\nabla\times E} & \,=-\frac{\partial\boldsymbol{B}}{\partial t}\,,\label{eq:38}\\
\boldsymbol{\nabla\times B}\, & =\,\frac{\partial\boldsymbol{E}}{\partial t}+\boldsymbol{J}\,.\label{eq:39}
\end{align}
Interestingly, if we combine together the conducting fluidic field
and electromagnetic field then the relevant theory comes out called
MHD. The MHD of cold plasma is an approximation theory of fluid dynamics
where we neglect temperature effect and combine the electron equation
with ionic equation to form a one-fluid model \cite{key-39}. For
the cold plasma model, many researchers \cite{key-40,key-41} suggested
that at a given position, all particle-species (mostly ions and electrons)
have comparable temperatures ($T$), energies ($\mathscr{E}$) (equivalent
to masses) and velocities ($\boldsymbol{v}$). It follows that the
fluid velocity is identical for particle velocity. Now, we may summarize
the following conditions for the cold plasma approximation, i.e.,
\begin{align}
T_{e}\,\,\, & \sim\,\,\,T_{i}\,\,\,\,(\text{neglected)}\nonumber \\
\mathscr{E}_{e}\,\,\, & \sim\,\,\,\mathscr{E}_{i}\,\nonumber \\
\boldsymbol{v}_{e}\,\,\, & \sim\,\,\,\boldsymbol{v}_{i}\,\nonumber \\
\rho_{e}\,\,\, & \sim\,\,\,\rho_{i}\,\nonumber \\
\boldsymbol{\nabla}p\,\,\, & \sim\,\,\,0\,.\label{eq:40}
\end{align}
We consider that the effected behavior of electrons are comparable
to the ions, while their temperatures and pressure-gradients are taken
negligible in case of homogeneous cold plasmas. Thus, using approximation
(\ref{eq:40}), the average mass and charge densities to cold plasma
are expressed as
\begin{align}
\varrho\,\longmapsto\,\rho_{M}(x)\,\,\simeq\, & \,\,m_{e}n_{e}(x)\,\,\equiv\,\,m_{i}n_{i}(x)\,,\label{eq:41}\\
\rho\,\longmapsto\,\rho_{c}(x)\,\,\simeq\, & \,\,q_{e}n_{e}(x)\,\,\equiv\,\,q_{i}n_{i}(x)\,.\label{eq:42}
\end{align}
As such, the Navier-Stokes and Ohm's equations become 
\begin{align}
\varrho\left(\frac{\partial}{\partial t}+\boldsymbol{v}\cdot\boldsymbol{\nabla}\right)\boldsymbol{v} & \,=\,\,\rho\boldsymbol{E}\,,\label{eq:43}\\
\boldsymbol{J}\,\,=\,\sigma(\boldsymbol{E}+ & \boldsymbol{v}\times\boldsymbol{B})\,,\label{eq:44}
\end{align}
where $\left(\boldsymbol{J\times B}\right)\sim0$ to the case if the
current is small compared to $\left(\boldsymbol{v}\times\boldsymbol{B}\right)$.
The ideal MHD equations ($\rho\sim0$) for cold plasma may then be
expressed as
\begin{align}
\frac{\partial\varrho}{\partial t}+\boldsymbol{\nabla}\cdot(\varrho\boldsymbol{v})\, & =\,\,0\,,\label{eq:45}\\
\varrho\left(\frac{\partial}{\partial t}+\boldsymbol{v}\cdot\boldsymbol{\nabla}\right)\boldsymbol{v}\, & =\,\,0\,,\label{eq:46}\\
\boldsymbol{\nabla}\times\left(\boldsymbol{v}\times\boldsymbol{B}\right)\, & =\,\,\frac{\partial\boldsymbol{B}}{\partial t}\,,\label{eq:47}\\
\boldsymbol{\nabla\times B}\, & =\,\,\frac{\partial\boldsymbol{E}}{\partial t}+\boldsymbol{J}\,.\label{eq:48}
\end{align}
To considering wave behavior of cold particles, the cold plasma wave
has temperature independent dispersion relation. If $\boldsymbol{v}_{A}$
is Alfven velocity, then the dispersion relation for cold plasma waves
become \cite{key-35} $\omega^{2}=\frac{\kappa^{2}\boldsymbol{v}_{A}^{2}}{1+\boldsymbol{v}_{A}^{2}}\,.$
Interestingly, the cold plasma waves propagate like as Alfven waves
which are independent on temperature.

\section{Dual MHD equations for dyonic cold plasma}

The dual MHD field consists not only electrons and ions but also having
the magnetic monopole and their ionic partners magneto-ions \cite{key-42}.
Generally, the composition of an electron and a magnetic monopole
referred a dyon \cite{key-25}. In this study, we may neglect the
magneto-ionic contribution like ions to continue the dyonic cold plasma
approximations. Dirac \cite{key-8} proposed the symmetrized field
equations by postulating the existence of magnetic monopoles, i.e.,
\begin{align}
\boldsymbol{\nabla\cdot E} & \,=\,\,\rho^{e}\,,\label{eq:49}\\
\boldsymbol{\nabla\cdot B} & \,=\,\,\rho^{\mathfrak{m}}\,,\label{eq:50}\\
\boldsymbol{\nabla\times E} & \,=\,-\frac{\partial\boldsymbol{B}}{\partial t}-\boldsymbol{J}^{\mathfrak{m}}\,,\label{eq:51}\\
\boldsymbol{\nabla\times B} & \,=\,\frac{\partial\boldsymbol{E}}{\partial t}+\boldsymbol{J}^{e}\,.\label{eq:52}
\end{align}
In the above generalized Dirac Maxwell's equations, $\rho^{e}$ and
$\rho^{\mathfrak{m}}$ are the electric and magnetic charge densities
while $\boldsymbol{J}^{e}$ and $\boldsymbol{J}^{\mathfrak{m}}$ are
the corresponding current densities. To study the dyonic cold plasma
field, there are a couple of masses and charges species in presence
of dyons. Thus, the generalized dual densities (mass and charge densities)
may be expressed for one-fluid theory of dyonic cold plasma as
\begin{align}
\varrho^{D}(\varrho^{e},\,\varrho^{\mathfrak{m}})\,\longmapsto\, & \left(m^{e}n^{e}+m^{\mathfrak{m}}n^{\mathfrak{m}}\right)\,,\label{eq:53}\\
\rho^{D}(\rho^{e},\,\rho^{\mathfrak{m}})\,\longmapsto\, & \left(q^{e}n^{e}+q^{\mathfrak{m}}n^{\mathfrak{m}}\right)\,,\label{eq:54}
\end{align}
where $m^{\mathfrak{m}},\,n^{\mathfrak{m}},$ and $q^{\mathfrak{m}}$
are defined the mass, total number and charge of magnetic monopoles,
respectively. As such, we can express the center of mass velocity
of dyonic fluid in cold plasma as
\begin{align}
\boldsymbol{v}^{D}\,\,\simeq & \,\,\,\frac{1}{\varrho^{D}}\left(\boldsymbol{v}^{e}m^{e}n^{e}(x)+\boldsymbol{v}^{\mathfrak{m}}m^{\mathfrak{m}}n^{\mathfrak{m}}(x)\right)\,,\label{eq:55}
\end{align}
whereupon the dual current densities (electric and magnetic) are defined
by
\begin{align}
\boldsymbol{J}^{e}\,\,= & \,\,q^{e}n^{e}\boldsymbol{v}^{e}\,,\,\,\,\,\text{and}\,\,\,\,\,\boldsymbol{J}^{\mathfrak{m}}\,=\,\,q^{\mathfrak{m}}n^{\mathfrak{m}}\boldsymbol{v}^{\mathfrak{m}}\,.\label{eq:56}
\end{align}
The conservation laws for the dynamics of dyonic cold plasma can be
written as

\begin{align}
\frac{\partial\varrho^{D}}{\partial t}+\boldsymbol{\nabla}\cdot(\varrho^{D}\boldsymbol{v}^{D})\, & =\,\,0\,,\,\,\,\,\,\,\,(\text{dyons mass conservation law})\label{eq:57}\\
\frac{\partial\rho^{e}}{\partial t}+\boldsymbol{\nabla}\cdot\boldsymbol{J}^{e}\, & =\,\,0\,,\,\,\,\,\,\,\,(\text{electric charge conservation law})\label{eq:58}\\
\frac{\partial\rho^{\mathfrak{m}}}{\partial t}+\boldsymbol{\nabla}\cdot\boldsymbol{J}^{\mathfrak{m}}\, & =\,\,0\,,\,\,\,\,\,\,\,(\text{magnetic charge conservation law})\,.\label{eq:59}
\end{align}
The generalized Navier-Stokes force equation can also be exhibited
in presence of magnetic monopole, i.e.
\begin{align}
\varrho^{D}\left(\frac{\partial}{\partial t}+\boldsymbol{v}^{D}\cdot\boldsymbol{\nabla}\right)\boldsymbol{v}^{D} & \,=\,\,\left(\boldsymbol{J}^{e}\boldsymbol{\times B}\right)-\left(\boldsymbol{J}^{\mathfrak{m}}\boldsymbol{\times E}\right)+\rho^{e}\boldsymbol{E}+\rho^{\mathfrak{m}}\boldsymbol{B}-\left(\boldsymbol{\nabla}p\right)^{D}\,,\label{eq:60}
\end{align}
where the duality invariant Lorentz force equation for dyons is
\begin{align}
\boldsymbol{F}^{D}\,\,= & \,\,\rho^{e}\boldsymbol{E}+\left(\boldsymbol{J}^{e}\boldsymbol{\times B}\right)+\rho^{\mathfrak{m}}\boldsymbol{B}-\left(\boldsymbol{J}^{\mathfrak{m}}\boldsymbol{\times E}\right)\,\label{eq:61}
\end{align}
and the dyonic pressure gradient term $\left(\boldsymbol{\nabla}p\right)^{D}$
takes negligible to the case of cold plasma approximation. Conditionally,
if the influence of dyonic current is small then the force equation
can be written as
\begin{align}
\varrho^{D}\left(\frac{\partial}{\partial t}+\boldsymbol{v}^{D}\cdot\boldsymbol{\nabla}\right)\boldsymbol{v}^{D}\, & =\,\,\rho^{e}\boldsymbol{E}+\rho^{\mathfrak{m}}\boldsymbol{B}\,.\label{eq:62}
\end{align}
In the same way, the Ohm's law for the dyonic cold plasma is expressed
as
\begin{alignat}{1}
\boldsymbol{J}^{e}\,= & \,\,\sigma^{e}(\boldsymbol{E}+\boldsymbol{v}\times\boldsymbol{B})\,,\label{eq:63}\\
\boldsymbol{J}^{\mathfrak{m}}\,= & \,\,\sigma^{\mathfrak{m}}(\boldsymbol{B}-\boldsymbol{v}\times\boldsymbol{E})\,.\label{eq:64}
\end{alignat}
where $\sigma^{\mathfrak{m}}$ is the magnetic conductivity. Therefore,
from equations (\ref{eq:63})-(\ref{eq:64}), we can conclude that
for infinite conductivity of dyons ($\sigma^{e,\mathfrak{m}}\rightarrow\infty$)
the electric and magnetic field vectors constitute from the rotation
of each other, i.e., $\boldsymbol{E}=-\left(\boldsymbol{v}\times\boldsymbol{B}\right),$
and $\boldsymbol{B}=\left(\boldsymbol{v}\times\boldsymbol{E}\right)$.
The above classical field equations given by (\ref{eq:49}) to (\ref{eq:64})
of dyons are referred to dual MHD field equations of dyonic cold plasma.

\section{Quaternionic formulation to dual fields of dyonic cold plasma}

In order to write the dual MHD field equations for dyonic cold plasma,
we may start with quaternionic two-velocity ($\boldsymbol{u},\,\boldsymbol{\upsilon}$)
and two-enthalpy ($h,\,k$) of dyons for plasma fluid dynamics as
\begin{align}
\mathbb{U}\left(e_{1},\,e_{2},\,e_{3},\,e_{0}\right)\,= & \,\left\{ u_{x},\,u_{y},\,u_{z},-\frac{i}{a_{0}}h\right\} \,,\label{eq:65}\\
\mathbb{V}\left(e_{1},\,e_{2},\,e_{3},\,e_{0}\right)\,= & \,\left\{ \upsilon_{x},\,\upsilon_{y},\,\upsilon_{z},-ia_{0}k\right\} \,,\label{eq:66}
\end{align}
where ($\mathbb{U},\,\mathbb{V}$) are quaternionic variables associated
with two four-velocities of electrons and magnetic monopoles of dyons
and, $a_{0}$ denoted the speed of particles (dyons) moving in conducting
cold plasma. Here, we have taken the two-enthalpy of dyons i.e. the
internal energy of dyons associated with electrons and magnetic monopoles.
Like many physicists \cite{key-32,key-43,key-44}, there is an analogy
between the electromagnetic and hydrodynamic. Thus, we may write the
analogy of two four-potentials ($\mathbb{A},\,\mathbb{B}$) of dyons
as
\begin{alignat}{1}
\mathbb{A}\left(\boldsymbol{\mathcal{A}},-\frac{i}{c}\phi^{e}\right)\,\,\,\longmapsto & \,\,\,\,\mathcal{\mathbb{U}}\left(\boldsymbol{u},-\frac{i}{a_{0}}h\right)\,,\label{eq:67}\\
\mathbb{B}\left(\boldsymbol{\mathcal{B}},-ic\phi^{\mathfrak{m}}\right)\,\,\,\longmapsto & \,\,\,\,\mathcal{\mathbb{V}}\left(\boldsymbol{\upsilon},-ia_{0}k\right)\,,\label{eq:68}
\end{alignat}
where the vector components $\boldsymbol{u}\rightarrow\left(u_{x},\,u_{y},\,u_{z}\right),\,\boldsymbol{\upsilon}\rightarrow\left(\upsilon_{x},\,\upsilon_{y},\,\upsilon_{z}\right)$
are analogous to electric and magnetic vector potentials of dyons
while the scalar components ($h$, $k$) are analogous to their scalar
potentials. It should be notice that the role of quaternionic two
four-velocities of dyonic-fluid in generalized hydrodynamics of cold
plasma is similar as the quaternionic two four-potentials of dyons
in generalized electrodynamics. Now, we may summarize the dyonic potentials
corresponding to its fluid behavior in table-1.\\
\begin{table}[H]
\begin{centering}
\begin{tabular}{cc}
\hline 
\textbf{Electrodynamics case} & \textbf{Hydrodynamics case}\tabularnewline
\hline 
\hline 
$\boldsymbol{\mathcal{A}}\,$ (electric vector potential)~~~ $\longmapsto$ & $\boldsymbol{u}\,$ (electric velocity of the fluid)\tabularnewline
$\boldsymbol{\mathcal{B}}\,$ (magnetic vector potential) ~~$\longmapsto$ & $\boldsymbol{\upsilon}\,$ (magnetic velocity of the fluid)\tabularnewline
$\phi^{e}\,$ (electric scalar potential) ~~~ $\longmapsto$ & $h\,$ (electric enthalpy of the fluid)\tabularnewline
$\phi^{\mathfrak{m}}\,$ (magnetic scalar potential) ~~$\longmapsto$ & $k\,$ (magnetic enthalpy of the fluid)\tabularnewline
\hline 
\end{tabular}
\par\end{centering}
\caption{Analogies between electrodynamics and hydrodynamics in presence of
dyons }
\end{table}
The unified structure of quaternionic two four-velocities ($\mathbb{W}\in\mathbb{H}$)
for the generalized fields of dyonic cold plasma can be written as
\begin{align}
\mathbb{W}\,= & \,\,\left(\mathbb{U}-\frac{i}{a_{0}}\mathbb{V}\right)\nonumber \\
= & \,\,\,e_{1}\left(u_{x}-\frac{i}{a_{0}}\upsilon_{x}\right)+e_{2}\left(u_{y}-\frac{i}{a_{0}}\upsilon_{y}\right)+e_{3}\left(u_{z}-\frac{i}{a_{0}}\upsilon_{z}\right)-\frac{i}{a_{0}}e_{0}(h-ia_{0}k)\,,\label{eq:69}
\end{align}
it reduces to
\begin{alignat}{1}
\mathbb{W}\,= & \,\,\sum_{j=1}^{3}e_{j}w_{j}-\frac{i}{a_{0}}e_{0}\Omega_{0}\nonumber \\
= & \,\,\sum_{j=1}^{3}e_{j}\left(u_{j}-\frac{i}{a_{0}}\upsilon_{j}\right)-\frac{i}{a_{0}}e_{0}(h-ia_{0}k)\,,\label{eq:70}
\end{alignat}
where $\boldsymbol{w}\rightarrow\left(\boldsymbol{u}-\frac{i}{a_{0}}\boldsymbol{\upsilon}\right)$
and $\Omega_{0}\rightarrow\left(h-ia_{0}k\right)$ are dyonic fluid-velocity
and dyonic enthalpy in cold plasma, respectively. Here, the scalar
component ($\Omega_{0}$) represents the amount of dyonic internal
energy required to move one kilogram of the fluid element. Now, to
formulate the quaternionic dual MHD field equations for dyonic cold
plasma, it is necessary to define quaternionic space-time differential
operator as

\begin{align}
\mathrm{\mathbb{D\,=}}\,\left(\boldsymbol{\nabla},\,-\frac{i}{a_{0}}\frac{\partial}{\partial t}\right)\,\simeq & \,\,\,\,e_{\mathrm{1}}\frac{\partial}{\partial x}+e_{2}\frac{\partial}{\partial y}+e_{3}\frac{\partial}{\partial z}-\frac{i}{a}e_{0}\frac{\partial}{\partial t}\,,\label{eq:71}
\end{align}
its quaternionic conjugate is

\begin{align}
\bar{\mathbb{D}}\,\,\mathrm{\mathbb{=}}\,\left(-\boldsymbol{\nabla},\,-\frac{i}{a_{0}}\frac{\partial}{\partial t}\right)\,\simeq & \,\,-e_{\mathrm{1}}\frac{\partial}{\partial x}-e_{2}\frac{\partial}{\partial y}-e_{3}\frac{\partial}{\partial z}-\frac{i}{a}e_{0}\frac{\partial}{\partial t}\,.\label{eq:72}
\end{align}
The quaternionic product of $\mathbb{D\circ\bar{D}}$ will be
\begin{alignat}{1}
\mathbb{D\circ\bar{D}}\,\,\,= & \,\,\,\frac{\partial^{2}}{\partial x^{2}}+\frac{\partial^{2}}{\partial y^{2}}+\frac{\partial^{2}}{\partial z^{2}}-\frac{1}{a_{0}^{2}}\frac{\partial^{2}}{\partial t^{2}}\nonumber \\
= & \,\,\,\boldsymbol{\nabla}^{2}-\frac{1}{a_{0}^{2}}\frac{\partial^{2}}{\partial t^{2}}\,\,=\,\,\mathbb{\bar{D}\circ D}\,,\label{eq:73}
\end{alignat}
where $\mathbb{D\circ\bar{D}}\,\,\text{or\,}\,\mathbb{\bar{D}\circ D}$
is defined by the D' Alembert operator $\square$. In order to emphasize
the variation of quaternionic space-time to two four-velocities of
dyonic fluid plasma, we may operate the quaternionic differential
operator ($\mathbb{D}$) on generalized two four-velocities ($\mathbb{W}$)
as

\begin{alignat}{1}
\mathbb{D\circ W}\,\,=\,\, & e_{1}\left\{ \left(\frac{\partial u_{z}}{\partial y}-\frac{\partial u_{y}}{\partial z}-\frac{1}{a_{0}^{2}}\frac{\partial\upsilon_{x}}{\partial t}-\frac{\partial k}{\partial x}\right)+\frac{i}{a_{0}}\left(-\frac{\partial\upsilon_{z}}{\partial y}+\frac{\partial\upsilon_{y}}{\partial z}-\frac{\partial u_{x}}{\partial t}-\frac{\partial h}{\partial x}\right)\right\} \nonumber \\
+ & e_{2}\left\{ \left(\frac{\partial u_{x}}{\partial z}-\frac{\partial u_{z}}{\partial x}-\frac{1}{a_{0}^{2}}\frac{\partial\upsilon_{y}}{\partial t}-\frac{\partial k}{\partial y}\right)+\frac{i}{a_{0}}\left(-\frac{\partial\upsilon_{x}}{\partial z}+\frac{\partial\upsilon_{z}}{\partial x}-\frac{\partial u_{y}}{\partial t}-\frac{\partial h}{\partial y}\right)\right\} \nonumber \\
+ & e_{3}\left\{ \left(\frac{\partial u_{y}}{\partial x}-\frac{\partial u_{x}}{\partial y}-\frac{1}{a_{0}^{2}}\frac{\partial\upsilon_{z}}{\partial t}-\frac{\partial k}{\partial z}\right)+\frac{i}{a_{0}}\left(-\frac{\partial\upsilon_{y}}{\partial x}+\frac{\partial\upsilon_{x}}{\partial y}-\frac{\partial u_{z}}{\partial t}-\frac{\partial h}{\partial z}\right)\right\} \nonumber \\
- & e_{0}\left\{ \left(\frac{\partial u_{x}}{\partial x}+\frac{\partial u_{y}}{\partial y}+\frac{\partial u_{z}}{\partial z}+\frac{1}{a_{0}^{2}}\frac{\partial h}{\partial t}\right)-\frac{i}{a_{0}}\left(\frac{\partial\upsilon_{x}}{\partial x}+\frac{\partial\upsilon_{y}}{\partial y}+\frac{\partial\upsilon_{z}}{\partial z}+\frac{\partial k}{\partial t}\right)\right\} \,\,.\label{eq:74}
\end{alignat}
Equation (\ref{eq:74}) governed the following quaternionic hydrodynamics
field equation for dyonic cold plasma, i.e.,
\begin{align}
\mathbb{D\,\circ W}\,\,=\,\,\boldsymbol{\Psi}\,\,\simeq & \,\,\,e_{1}\psi_{1}+e_{2}\psi_{2}+e_{3}\psi_{3}+e_{0}\chi\,,\label{eq:75}
\end{align}
where $\boldsymbol{\psi}\rightarrow(\psi_{1},\,\psi_{2},\,\psi_{3})$
and $\chi$ are the vector and scalar fields connected to the hydrodynamics
of dyonic cold plasma, respectively. Further, the unified structure
of quaternionic hydrodynamics field components can be expressed as
\begin{align}
\psi_{1}\,= & \,\left\{ \left(\boldsymbol{\nabla}\times\boldsymbol{u}\right)_{x}-\frac{1}{a_{0}^{2}}\frac{\partial\upsilon_{x}}{\partial t}-\frac{\partial k}{\partial x}\right\} +\frac{i}{a_{0}}\left\{ -\left(\boldsymbol{\nabla}\times\boldsymbol{\upsilon}\right)_{x}-\frac{\partial u_{x}}{\partial t}-\frac{\partial h}{\partial x}\right\} \,,\label{eq:76}\\
\psi_{2}\,= & \,\left\{ \left(\boldsymbol{\nabla}\times\boldsymbol{u}\right)_{y}-\frac{1}{a_{0}^{2}}\frac{\partial\upsilon_{y}}{\partial t}-\frac{\partial k}{\partial y}\right\} +\frac{i}{a_{0}}\left\{ -\left(\boldsymbol{\nabla}\times\boldsymbol{\upsilon}\right)_{y}-\frac{\partial u_{y}}{\partial t}-\frac{\partial h}{\partial y}\right\} \,,\label{eq:77}\\
\psi_{3}\,= & \,\left\{ \left(\boldsymbol{\nabla}\times\boldsymbol{u}\right)_{z}-\frac{1}{a_{0}^{2}}\frac{\partial\upsilon_{z}}{\partial t}-\frac{\partial k}{\partial z}\right\} +\frac{i}{a_{0}}\left\{ -\left(\boldsymbol{\nabla}\times\boldsymbol{\upsilon}\right)_{z}-\frac{\partial u_{z}}{\partial t}-\frac{\partial h}{\partial z}\right\} \,,\label{eq:78}\\
\chi\,= & -\left\{ \left(\boldsymbol{\nabla}\cdotp\boldsymbol{u}+\frac{1}{a_{0}^{2}}\frac{\partial h}{\partial t}\right)-\frac{i}{a_{0}}\left(\boldsymbol{\nabla}\cdotp\boldsymbol{\upsilon}+\frac{\partial k}{\partial t}\right)\right\} \,.\label{eq:79}
\end{align}
We may consider the generalized dual hydrodynamics fields namely the
hydro-electric and hydro-magnetic fields of dyonic-fluid associated
with the dynamics of electrons and magnetic monopoles in dyonic cold
plasma. Thus, the unified fields can be rewrite as
\begin{align}
\psi_{1} & \,\,\longleftrightarrow\,\,\left(B_{x}+\frac{i}{a_{0}}E_{x}\right)\,,\label{eq:80}\\
\psi_{2} & \,\,\longleftrightarrow\,\,\left(B_{y}+\frac{i}{a_{0}}E_{y}\right)\,,\label{eq:81}\\
\psi_{2} & \,\,\longleftrightarrow\,\,\left(B_{y}+\frac{i}{a_{0}}E_{y}\right)\,,\label{eq:82}\\
\chi & \,\,\longleftrightarrow\,-\left(\mathcal{L}-\frac{i}{a_{0}}\widetilde{\mathcal{L}}\right)\,.\label{eq:83}
\end{align}
The hydro-electric field vector ($\boldsymbol{E}$) plays as the generalized
Lamb vector field and the hydro-magnetic field vector ($\boldsymbol{B}$)
plays as the generalized vorticity field \cite{key-45,key-46,key-47}
to the case of dual MHD. The generalized Lamb vector field may be
used to accelerate the dyonic fluid flow while the vorticity field
is its counterpart. Thus, the generalized dual fields ($\boldsymbol{E},\,\boldsymbol{B}$)
for dyonic fluid become,
\begin{align}
\boldsymbol{E}\,= & \,-\boldsymbol{\nabla}\times\boldsymbol{\upsilon}-\frac{\partial\boldsymbol{u}}{\partial t}-\boldsymbol{\nabla}h\,,\label{eq:84}\\
\boldsymbol{B}\,= & \,\,\boldsymbol{\nabla}\times\boldsymbol{u}-\frac{1}{a_{0}^{2}}\frac{\partial\boldsymbol{\upsilon}}{\partial t}-\boldsymbol{\nabla}k\,,\label{eq:85}
\end{align}
and the dual Lorenz gauge conditions ($\mathcal{L}$, $\widetilde{\mathcal{L}}$)
for the continuous flow of incompressible dyonic fluid plasma are
\begin{alignat}{1}
\mathcal{L}\,\,:\,\longmapsto\,\,\,\, & \boldsymbol{\nabla}\cdotp\boldsymbol{u}+\frac{1}{a_{0}^{2}}\frac{\partial h}{\partial t}\,=\,\,0\,,\label{eq:86}\\
\widetilde{\mathcal{L}}\,\,:\,\longmapsto\,\,\,\, & \boldsymbol{\nabla}\cdotp\boldsymbol{\upsilon}+\frac{\partial k}{\partial t}\,=\,\,0\,.\label{eq:87}
\end{alignat}
The unified quaternionic Lamb-vorticity field vector $\boldsymbol{\Psi}$
(or generalized hydro-electromagnetic field vector) for dyons can
be expressed as
\begin{alignat}{1}
\boldsymbol{\Psi}\,\, & =\,\,e_{1}\left(B_{x}+\frac{i}{a_{0}}E_{x}\right)+e_{2}\left(B_{y}+\frac{i}{a_{0}}E_{y}\right)+e_{3}\left(B_{z}+\frac{i}{a_{0}}E_{z}\right)\,.\label{eq:88}
\end{alignat}
Now, applying the quaternionic conjugate of differential operator
$\mathbb{\bar{D}}$ to equation (\ref{eq:88}), we obtain

\begin{alignat}{1}
\bar{\mathbb{D}}\circ\boldsymbol{\Psi}\,\,=\,\,- & \,e_{1}\left[\left\{ \left(\boldsymbol{\nabla}\times\boldsymbol{B}\right)_{x}-\frac{1}{a_{0}^{2}}\frac{\partial E_{x}}{\partial t}\right\} +\frac{i}{a_{0}}\left\{ \left(\boldsymbol{\nabla}\times\boldsymbol{E}\right)_{x}+\frac{\partial B_{x}}{\partial t}\right\} \right]\nonumber \\
- & \,e_{2}\left[\left\{ \left(\boldsymbol{\nabla}\times\boldsymbol{B}\right)_{y}-\frac{1}{a_{0}^{2}}\frac{\partial E_{y}}{\partial t}\right\} +\frac{i}{a_{0}}\left\{ \left(\boldsymbol{\nabla}\times\boldsymbol{E}\right)_{y}+\frac{\partial B_{y}}{\partial t}\right\} \right]\nonumber \\
- & \,e_{3}\left[\left\{ \left(\boldsymbol{\nabla}\times\boldsymbol{B}\right)_{z}-\frac{1}{a_{0}^{2}}\frac{\partial E_{z}}{\partial t}\right\} +\frac{i}{a_{0}}\left\{ \left(\boldsymbol{\nabla}\times\boldsymbol{E}\right)_{z}+\frac{\partial B_{z}}{\partial t}\right\} \right]\nonumber \\
+ & \,e_{0}\left[\boldsymbol{\nabla}\cdotp\boldsymbol{B}+\frac{i}{a_{0}}\boldsymbol{\nabla}\cdotp\boldsymbol{E}\right]\,.\label{eq:89}
\end{alignat}
Equation (\ref{eq:89}) shows the quaternionic space-time evaluation
of generalized Lamb-vorticity fields in the incompressible fluid of
dyonic cold plasma. The dynamics of dyonic cold plasma fluid can be
expressed by following equation
\begin{alignat}{1}
\bar{\mathbb{D}}\circ\boldsymbol{\Psi}\,\,= & \,-\mathbb{S}\,(\boldsymbol{S},\,\text{\ensuremath{\wp}})\,\,\simeq\,\,-\left(e_{1}S_{1}+e_{2}S_{2}+e_{3}S_{3}+e_{0}\text{\ensuremath{\wp}}\right)\,,\label{eq:90}
\end{alignat}
where $\mathbb{S}$ is the quaternionic source for the dyonic cold
plasma. Moreover, the quaternionic vector and scalar components of
dyonic sources, i.e., ($\boldsymbol{S},\,\text{\ensuremath{\wp}}$)
can be written as
\begin{align}
S_{1}\,\,\longleftrightarrow\,\, & \left(\mu J_{x}^{e}-\frac{i}{a_{0}}\frac{J_{x}^{\mathfrak{m}}}{\epsilon}\right)\,,\label{eq:91}\\
S_{2}\,\,\longleftrightarrow\,\, & \left(\mu J_{y}^{e}-\frac{i}{a_{0}}\frac{J_{y}^{\mathfrak{m}}}{\epsilon}\right)\,,\label{eq:92}\\
S_{3}\,\,\longleftrightarrow\,\, & \left(\mu J_{z}^{e}-\frac{i}{a_{0}}\frac{J_{z}^{\mathfrak{m}}}{\epsilon}\right)\,,\label{eq:93}\\
\text{\ensuremath{\wp}}\,\,\longleftrightarrow\,\, & \left(\mu\rho^{\mathfrak{m}}-\frac{i}{a_{0}}\frac{\rho^{e}}{\epsilon}\right)\,,\label{eq:94}
\end{align}
where ($\boldsymbol{J}^{e}$, $\rho^{e}$) are the quaternionic electric
source current and source density associated with the dynamics of
hydro-electric field while ($\boldsymbol{J}^{\mathfrak{m}}$,$\rho^{\mathfrak{m}}$)
are corresponding magnetic sources associated with the dynamics of
hydro-magnetic field of dyonic fluid. Therefore, the quaternionic
unified hydro-electromagnetic source for dyonic cold plasma can be
expressed by
\begin{alignat}{1}
\mathbb{S}\,\,= & \,\,\mu\left(e_{1}J_{x}^{e}+e_{2}J_{y}^{e}+e_{3}J_{z}^{e}-e_{0}\rho^{\mathfrak{m}}\right)-\frac{i}{a_{0}}\left(e_{1}\frac{J_{x}^{\mathfrak{m}}}{\epsilon}+e_{2}\frac{J_{y}^{\mathfrak{m}}}{\epsilon}+e_{3}\frac{J_{z}^{\mathfrak{m}}}{\epsilon}+e_{0}\frac{\rho^{e}}{\epsilon}\right)\nonumber \\
= & \,\,\left(\mu e_{j}\boldsymbol{J}^{e}-\frac{ie_{0}}{a_{0}}\frac{\rho^{e}}{\epsilon}\right)-\frac{i}{a_{0}}\left(e_{j}\frac{\boldsymbol{J}^{\mathfrak{m}}}{\epsilon}-ie_{0}a_{0}\mu\rho^{\mathfrak{m}}\right)\nonumber \\
= & \,\,\left(\mathbb{J}-\frac{i}{a_{0}}\mathbb{K}\right)\,.\label{eq:95}
\end{alignat}
Here, $\mathbb{J}(e_{j},\,e_{0})\rightarrow\left(\mu\boldsymbol{J}^{e}\,,\,\,-\frac{i}{a_{0}}\frac{\rho^{e}}{\epsilon}\right),\,\mathbb{K}(e_{j},\,e_{0})\rightarrow\left(\frac{1}{\epsilon}\boldsymbol{J}^{\mathfrak{m}}\,,\,\,-ia_{0}\mu\rho^{\mathfrak{m}}\right)$
are quaternionic two four-fluid sources of dyons and ($\epsilon,\,\mu$)
are considering the permittivity and permeability satisfy $a_{0}=\frac{1}{\sqrt{\mu\epsilon}}$.
Now, equating quaternionic imaginary and real coefficients in equation
(\ref{eq:90}), and obtain,
\begin{align}
\boldsymbol{\nabla}\cdotp\boldsymbol{E}\,\,= & \,\,\frac{\rho^{e}}{\epsilon}\,\,,\,\,\,\,\,\,\,\,\,\,\,\,\,\,\,\,\,\,\,\,\,\,\text{(Imaginary part of }e_{0})\label{eq:96}\\
\boldsymbol{\nabla}\cdotp\boldsymbol{B}\,\,= & \,\,\mu\rho^{\mathfrak{m}}\,\,,\,\,\,\,\,\,\,\,\,\,\,\,\,\,\,\,\,\,\,\,\text{(Real part of \ensuremath{e_{0}}})\label{eq:97}\\
\left(\boldsymbol{\nabla}\times\boldsymbol{E}\right)_{x}\,= & \,-\frac{\partial B_{x}}{\partial t}-\frac{J_{x}^{\mathfrak{m}}}{\epsilon}\,\,,\,\,\,\,\,\text{\,\,\,\,\,\,(Imaginary part of }e_{1})\label{eq:98}\\
\left(\boldsymbol{\nabla}\times\boldsymbol{E}\right)_{y}\,= & \,-\frac{\partial B_{y}}{\partial t}-\frac{J_{y}^{\mathfrak{m}}}{\epsilon}\,\,,\,\,\,\,\,\,\,\,\,\,\,\text{(Imaginary part of }e_{2})\label{eq:99}\\
\left(\boldsymbol{\nabla}\times\boldsymbol{E}\right)_{z}\,= & \,-\frac{\partial B_{z}}{\partial t}-\frac{J_{z}^{\mathfrak{m}}}{\epsilon}\,\,,\,\,\,\,\,\,\,\,\,\,\,\text{(Imaginary part of }e_{3})\label{eq:100}\\
\left(\boldsymbol{\nabla}\times\boldsymbol{B}\right)_{x}\,= & \,\,\frac{1}{a_{0}^{2}}\frac{\partial E_{x}}{\partial t}+\mu J_{x}^{e}\,\,,\,\,\,\,\:\,\,\,\,\text{(Real part of }e_{1})\label{eq:101}\\
\left(\boldsymbol{\nabla}\times\boldsymbol{B}\right)_{y}\,= & \,\,\frac{1}{a_{0}^{2}}\frac{\partial E_{y}}{\partial t}+\mu J_{y}^{e}\,\,,\,\,\,\,\,\,\,\,\,\,\text{(Real part of }e_{2})\label{eq:102}\\
\left(\boldsymbol{\nabla}\times\boldsymbol{B}\right)_{z}\,= & \,\,\frac{1}{a_{0}^{2}}\frac{\partial E_{z}}{\partial t}+\mu J_{z}^{e}\,\,,\,\,\,\,\,\,\,\,\,\,\,\text{(Real part of }e_{3})\,.\label{eq:103}
\end{align}
The above eight equations represent the quaternionic field equations
for hydrodynamics of dyonic cold plasma. These obtained equations
are primary equations for dual MHD of dyonic cold plasma, which are
exactly same as the generalized Dirac-Maxwell equations given by (\ref{eq:49})-(\ref{eq:52}).
As such, we also may write the unified dual MHD field equations for
dyonic cold plasma as

\begin{alignat}{1}
\boldsymbol{\nabla}\cdotp\boldsymbol{\Psi}\,\,= & \,\,i\text{\ensuremath{\wp}}\,,\label{eq:104}\\
\boldsymbol{\nabla}\times\boldsymbol{\Psi}\,\,= & \,-\frac{i}{a_{0}}\frac{\partial\boldsymbol{\Psi}}{\partial t}+\boldsymbol{S}\,.\label{eq:105}
\end{alignat}
The present quaternionic formulation describes the macroscopic cold
plasma behavior. The solution of differential equations (\ref{eq:104})-(\ref{eq:105})
provide the evolution of generalized lamb vector field and generalized
vorticity field to the presence of dyonic cold plasma. Now, we may
check the validity of dual MHD field equations for dyonic cold plasma
in given subsections.

\subsection{Duality invariant}

Let us check the duality invariant symmetry for generalized hydro-electric
and hydro-magnetic fields of dyonic cold plasma. The duality transformation
defines the rotation of hydro-electric and hydro-magnetic field components
in the quaternionic space such that the physics behind the quantity
remains the same after the transformation is performed. Suppose, $F^{\alpha\beta}$
and $\mathcal{F}^{\alpha\beta}$ are the field and dual field tensor,
then the duality transformation becomes \cite{key-48}
\begin{alignat}{1}
F^{'\alpha\beta}\,:\,\,\longmapsto\,\, & F^{\alpha\beta}\,\cos\theta+\mathcal{F^{\alpha\beta}}\,\sin\theta\,,\nonumber \\
\mathcal{F}^{'\alpha\beta}\,:\,\,\longmapsto\,\, & -\mathcal{F}^{\alpha\beta}\,\sin\theta+F^{\alpha\beta}\,\cos\theta\,,\,\,\,\,\,\,\,\,\,\,\,\,\,\,(0\leq\theta\leq\frac{\pi}{2})\,.\label{eq:106}
\end{alignat}
Correspondingly, the quaternionic hydro-electric and hydro-magnetic
fields can also transform as
\begin{alignat}{1}
\begin{pmatrix}\boldsymbol{E}\\
\boldsymbol{B}
\end{pmatrix}\,\,\longmapsto\,\,\mathfrak{D}_{2\times2} & \begin{pmatrix}\boldsymbol{E}\\
\boldsymbol{B}
\end{pmatrix}\,,\label{eq:107}
\end{alignat}
where $\mathfrak{D}_{2\times2}=\begin{pmatrix}\cos\theta & a_{0}\sin\theta\\
-\frac{1}{a_{0}}\sin\theta & \cos\theta
\end{pmatrix}$ is an unitary matrix called the duality transformation matrix (or
simply D-matrix). For general case $\theta=\frac{\pi}{2}$, the generalized
dual fields will be transform as
\begin{alignat}{1}
\begin{pmatrix}\boldsymbol{E}\\
\boldsymbol{B}
\end{pmatrix}\,\,=\,\,\begin{pmatrix}0 & a_{0}\\
-\frac{1}{a_{0}} & 0
\end{pmatrix} & \begin{pmatrix}\boldsymbol{E}\\
\boldsymbol{B}
\end{pmatrix}\,:\,\,\Longrightarrow\,\,\,\begin{cases}
\boldsymbol{E}\longmapsto\,a_{0}\boldsymbol{B} & ,\\
\boldsymbol{B}\longmapsto\,-\frac{1}{a_{0}}\boldsymbol{E} & .
\end{cases}\label{eq:108}
\end{alignat}
Here, the D-matrix $\mathfrak{D}_{2\times2}=\begin{pmatrix}0 & a_{0}\\
-\frac{1}{a_{0}} & 0
\end{pmatrix}$. For quaternionic dual-velocity and dual-enthalpy of dyons fluid,
the following duality transformation relations governed the streamline
flow, i.e.,
\begin{align}
\begin{pmatrix}\boldsymbol{u}\\
\boldsymbol{\upsilon}
\end{pmatrix}\,\,=\,\,\begin{pmatrix}0 & a_{0}\\
-\frac{1}{a_{0}} & 0
\end{pmatrix} & \begin{pmatrix}\boldsymbol{u}\\
\boldsymbol{\upsilon}
\end{pmatrix}\,:\,\,\Longrightarrow\,\,\,\begin{cases}
\boldsymbol{u}\longmapsto\,a_{0}\boldsymbol{\upsilon} & ,\\
\boldsymbol{\upsilon}\longmapsto\,-\frac{1}{a_{0}}\boldsymbol{u} & ,
\end{cases}\label{eq:109}\\
\begin{pmatrix}h\\
k
\end{pmatrix}\,\,=\,\,\begin{pmatrix}0 & a_{0}\\
-\frac{1}{a_{0}} & 0
\end{pmatrix} & \begin{pmatrix}h\\
k
\end{pmatrix}\,:\,\,\Longrightarrow\,\,\,\begin{cases}
h\longmapsto\,a_{0}k & ,\\
k\longmapsto\,-\frac{1}{a_{0}}h & .
\end{cases}\label{eq:110}
\end{align}
Accordingly, the dual-current and dual-density of dyonic plasma will
be transform as
\begin{align}
\begin{pmatrix}\boldsymbol{J}^{e}\\
\boldsymbol{J}^{\mathfrak{m}}
\end{pmatrix}\,\,=\,\,\begin{pmatrix}0 & a_{0}\\
-\frac{1}{a_{0}} & 0
\end{pmatrix} & \begin{pmatrix}\boldsymbol{J}^{e}\\
\boldsymbol{J}^{\mathfrak{m}}
\end{pmatrix}\,:\,\,\Longrightarrow\,\,\,\begin{cases}
\boldsymbol{J}^{e}\longmapsto\,a_{0}\boldsymbol{J}^{\mathfrak{m}} & ,\\
\boldsymbol{J}^{\mathfrak{m}}\longmapsto\,-\frac{1}{a_{0}}\boldsymbol{J}^{e} & ,
\end{cases}\label{eq:111}\\
\begin{pmatrix}\rho^{e}\\
\rho^{\mathfrak{m}}
\end{pmatrix}\,\,=\,\,\begin{pmatrix}0 & a_{0}\\
-\frac{1}{a_{0}} & 0
\end{pmatrix} & \begin{pmatrix}\rho^{e}\\
\rho^{\mathfrak{m}}
\end{pmatrix}\,:\,\,\Longrightarrow\,\,\,\begin{cases}
\rho^{e}\longmapsto\,a_{0}\rho^{\mathfrak{m}} & ,\\
\rho^{\mathfrak{m}}\longmapsto\,-\frac{1}{a_{0}}\rho^{e} & .
\end{cases}\label{eq:112}
\end{align}
Interestingly, from relations (\ref{eq:108}) to (\ref{eq:112}),
we can conclude that the generalized Dirac- Maxwell equations for
dyonic fluid of cold plasma are invariant under the duality transformations
and showing the highly symmetric nature in presence of dyonic fluid.

\subsection{Lorentz invariant}

Let us start with the most usual transformation \textbf{\cite{key-49,key-50}}
that preserves the quaternionic intervals $ds^{2}=dx^{2}+dy^{2}+dz^{2}-a_{0}^{2}dt^{2}$,
i.e.,\textbf{
\begin{alignat}{1}
X^{'\xi}\,\,= & \,\,\Lambda_{\eta}^{\xi}X^{\eta}\,,\label{eq:113}
\end{alignat}
}where $X$ is any four-vector and the Lorentz transformation matrix
element $\Lambda_{\eta}^{\xi}$ is 
\begin{align}
\Lambda_{\eta}^{\xi}\,\,\longmapsto & \,\,\begin{pmatrix}\cosh\phi & 0 & 0 & -i\,\sinh\phi\\
0 & 1 & 0 & 0\\
0 & 0 & 1 & 0\\
i\,\sinh\phi & 0 & 0 & \cosh\phi
\end{pmatrix}\,.\label{eq:114}
\end{align}
Here $\phi$ is the boost parameter. Using the above Lorentz transformation
matrix, we may obtain the following transformation equations for quaternionic
four-velocity ($\mathbb{W}$) of dyonic cold plasma which are an analogous
to quaternionic potentials of dyons, i.e.,

\begin{alignat}{1}
w_{x}^{'} & \,=\,\,\gamma\left(w_{x}-a_{0}\Omega_{0}\right)\,,\,\,\,w_{y}^{'}\,=\,w_{y}\,,\,\,\,w_{z}^{'}\,=\,w_{z}\,,\nonumber \\
\Omega_{0}^{'} & \,=\,\,\gamma\left(\Omega_{0}-a_{0}w_{x}\right)\,,\label{eq:115}
\end{alignat}
where
\begin{alignat}{1}
\cosh\phi\,\,= & \,\,\,\frac{1}{\sqrt{1-\tanh{}^{2}\phi}}\,\,=\,\,\frac{1}{\sqrt{1-a_{0}^{2}}}\,=\,\gamma\,,\nonumber \\
\sinh\phi\,\,= & \,\,\,a_{0}\gamma\,.\label{eq:116}
\end{alignat}
If we consider the massive dyonic particles \cite{key-51}, then the
transformation relations (\ref{eq:115}) lead to the energy-momentum
transformations for dyonic cold plasma,
\begin{align}
\mathscr{P}_{x}^{'} & \,=\,\,\gamma\left(\mathscr{P}_{x}-a_{0}\mathscr{E}\right)\,,\,\,\,\,\mathscr{P}_{y}^{'}\,=\,\mathscr{P}_{y}\,,\,\,\,\mathscr{P}_{z}^{'}\,=\,\mathscr{P}_{z}\,,\nonumber \\
\mathscr{E}{}^{'} & \,=\,\,\gamma\left(\mathscr{E}-a_{0}\mathscr{P}_{x}\right)\,,\label{eq:117}
\end{align}
where the quaternionic four-momentum is defined by $\mathbb{P}\,(e_{1},\,e_{2},\,e_{3},\,e_{0})=(\mathscr{P}_{x},\,\mathscr{P}_{y},\,\mathscr{P}_{z},\,\mathscr{E})$.
It should be notice that the obtained relations (\ref{eq:117}) are
similar to the usual relativistic Lorentz energy-momentum transformation
relations \cite{key-49,key-50}, where we assume that the speed of
dyons ($a_{0}$) is comparable to the speed of light ($c\sim1$).
As such, we also may establish the following transformation relations
for quaternionic source current and source density, i.e.,

\begin{alignat}{1}
S_{x}^{'}\, & =\,\,\gamma\left(S_{x}-a_{0}\text{\ensuremath{\wp}}\right)\,,\,\,\,\,S_{y}^{'}\,=\,S_{y}\,,\,\,\,\,S_{z}^{'}\,=\,S_{z}\,,\nonumber \\
\text{\ensuremath{\wp}}^{'}\, & =\,\,\gamma\left(\text{\ensuremath{\wp}}-a_{0}S_{x}\right)\,.\label{eq:118}
\end{alignat}
Correspondingly, we obtain the Lorentz transformation relations for
unified hydro-electromagnetic field of dyonic cold plasma, so that,
\begin{alignat}{1}
\psi_{x}^{'}\,\,=\,\,\psi_{x}\,,\,\,\,\psi_{y}^{'}\,\,=\,\,\gamma\left(\psi_{y}-ia_{0}\psi_{z}\right),\,\,\,\psi_{z}^{'}\,\,= & \,\,\gamma\left(\psi_{z}+ia_{0}\psi_{y}\right)\,,\label{eq:119}
\end{alignat}
along with 
\begin{alignat}{1}
\frac{\partial}{\partial x^{'}}\,\,= & \,\,\gamma\left(\frac{\partial}{\partial x}+\frac{\partial}{\partial t}\right)\,,\,\,\,\,\,\,\frac{\partial}{\partial t^{'}}\,\,=\,\,\gamma\left(\frac{\partial}{\partial t}+a_{0}^{2}\frac{\partial}{\partial x}\right)\,.\label{eq:120}
\end{alignat}
The beauty of the transformation relations (\ref{eq:118})-(\ref{eq:120})
is that, the generalized Dirac-Maxwell equations for dyonic fluid
of cold plasma are well invariant under these Lorentz transformation.

\subsection{CPT Invariant}

In order to check the CPT invariance \cite{key-52} for the dual MHD
field equations of dyonic cold plasma, we may write the charge conjugation
matrix ($\mathcal{C}$) to the case of quaternionic dual-current sources
and hydro-electromagnetic fields of dyonic fluid as $\mathcal{C}\,\rightarrow\,\,\begin{pmatrix}-1 & 0\\
0 & -1
\end{pmatrix}$, where the charge conjugation transformation plays as
\begin{align}
\mathcal{C}:\,\,\,\,\,\,\,\,\,\,\,\begin{pmatrix}\boldsymbol{J}^{'e}\\
\boldsymbol{J}^{'\mathfrak{m}}
\end{pmatrix}\,\,\longmapsto & \,\,\begin{pmatrix}-1 & 0\\
0 & -1
\end{pmatrix}\begin{pmatrix}\boldsymbol{J}^{e}\\
\boldsymbol{J}^{\mathfrak{m}}
\end{pmatrix}\,,\label{eq:121}\\
\mathcal{C}:\,\,\,\,\,\,\,\,\,\,\,\,\,\begin{pmatrix}\boldsymbol{E}^{'}\\
\boldsymbol{B}^{'}
\end{pmatrix}\,\,\longmapsto & \,\,\begin{pmatrix}-1 & 0\\
0 & -1
\end{pmatrix}\begin{pmatrix}\boldsymbol{E}\\
\boldsymbol{B}
\end{pmatrix}\,.\label{eq:122}
\end{align}
Correspondingly, the parity matrix $P\rightarrow\begin{pmatrix}-1 & 0\\
0 & 1
\end{pmatrix}$ can govern the following transformations for the dyonic fluid,
\begin{align}
P:\,\,\,\,\,\,\,\,\,\,\,\begin{pmatrix}\boldsymbol{J}^{'e}\\
\boldsymbol{J}^{'\mathfrak{m}}
\end{pmatrix}\,\,\longmapsto & \,\,\begin{pmatrix}-1 & 0\\
0 & 1
\end{pmatrix}\begin{pmatrix}\boldsymbol{J}^{e}\\
\boldsymbol{J}^{\mathfrak{m}}
\end{pmatrix}\,,\label{eq:123}\\
P:\,\,\,\,\,\,\,\,\,\,\,\,\,\begin{pmatrix}\boldsymbol{E}^{'}\\
\boldsymbol{B}^{'}
\end{pmatrix}\,\,\longmapsto & \,\,\begin{pmatrix}-1 & 0\\
0 & 1
\end{pmatrix}\begin{pmatrix}\boldsymbol{E}\\
\boldsymbol{B}
\end{pmatrix}\,.\label{eq:124}
\end{align}
As such, we can write the time reversal matrix, i.e. $T\rightarrow\begin{pmatrix}1 & 0\\
0 & -1
\end{pmatrix}$, and the transformation perform as
\begin{align}
T:\,\,\,\,\,\,\,\,\,\,\,\begin{pmatrix}\boldsymbol{J}^{'\mathfrak{m}}\\
\boldsymbol{J}^{'e}
\end{pmatrix}\,\,\longmapsto & \,\,\begin{pmatrix}1 & 0\\
0 & -1
\end{pmatrix}\begin{pmatrix}\boldsymbol{J}^{\mathfrak{m}}\\
\boldsymbol{J}^{e}
\end{pmatrix}\,,\label{eq:125}\\
T:\,\,\,\,\,\,\,\,\,\,\,\,\,\begin{pmatrix}\boldsymbol{E}^{'}\\
\boldsymbol{B}^{'}
\end{pmatrix}\,\,\longmapsto & \,\,\begin{pmatrix}1 & 0\\
0 & -1
\end{pmatrix}\begin{pmatrix}\boldsymbol{E}\\
\boldsymbol{B}
\end{pmatrix}\,.\label{eq:126}
\end{align}
The forth component of quaternionic sources can also be transform
for charge conjugation, parity and time reversal as the following
ways
\begin{alignat}{1}
\mathcal{C}:\,\,\,\,\,\,\,\,\,\,\,\begin{pmatrix}\rho^{'e}\\
\rho^{'\mathfrak{m}}
\end{pmatrix}\,\,\longmapsto & \,\begin{pmatrix}-1 & 0\\
0 & 1
\end{pmatrix}\begin{pmatrix}\rho^{e}\\
\rho^{\mathfrak{m}}
\end{pmatrix}\,,\label{eq:127}\\
P:\,\,\,\,\,\,\,\,\,\,\,\begin{pmatrix}\rho^{'e}\\
\rho^{'\mathfrak{m}}
\end{pmatrix}\,\,\,\longmapsto & \,\,\begin{pmatrix}1 & 0\\
0 & -1
\end{pmatrix}\begin{pmatrix}\rho^{e}\\
\rho^{\mathfrak{m}}
\end{pmatrix}\,,\label{eq:128}\\
T:\,\,\,\,\,\,\,\,\,\,\,\begin{pmatrix}\rho^{'e}\\
\rho^{'\mathfrak{m}}
\end{pmatrix}\,\,\longmapsto & \,\,\begin{pmatrix}1 & 0\\
0 & 1
\end{pmatrix}\begin{pmatrix}\rho^{e}\\
\rho^{\mathfrak{m}}
\end{pmatrix}\,.\label{eq:129}
\end{alignat}
We can summarize the quaternionic physical quantities of dual MHD
fields and their changes under charge conjugation, parity inversion
and time reversal given by table-2 \cite{key-53,key-54}.\\
\begin{table}[H]
\begin{centering}
\begin{tabular}{cccc}
\hline 
\textbf{Physical quantities} & \textbf{Charge conjugation ($\mathcal{C}$)} & \textbf{Parity} \textbf{inversion ($P$)} & \textbf{Time reversal ($T$)}\tabularnewline
\hline 
\hline 
\multirow{1}{*}{$\partial_{t}$} & $\partial_{t}$ & $\partial_{t}$ & $-\partial_{t}$\tabularnewline
$\boldsymbol{\nabla}$ & $\boldsymbol{\nabla}$ & $-\boldsymbol{\nabla}$ & $\boldsymbol{\nabla}$\tabularnewline
$a_{0}$ & $a_{0}$ & $-a_{0}$ & $-a_{0}$\tabularnewline
$\boldsymbol{J}^{e}$ & $-\boldsymbol{J}^{e}$ & $-\boldsymbol{J}^{e}$ & $-\boldsymbol{J}^{e}$\tabularnewline
$\boldsymbol{J}^{\mathfrak{m}}$ & $-\boldsymbol{J}^{\mathfrak{m}}$ & $\boldsymbol{J}^{\mathfrak{m}}$ & $\boldsymbol{J}^{\mathfrak{m}}$\tabularnewline
$\boldsymbol{E}$ & $-\boldsymbol{E}$ & $-\boldsymbol{E}$ & $\boldsymbol{E}$\tabularnewline
$\boldsymbol{B}$ & $-\boldsymbol{B}$ & $\boldsymbol{B}$ & $-\boldsymbol{B}$\tabularnewline
$\rho^{e}$ & $-\rho^{e}$ & $\rho^{e}$ & $\rho^{e}$\tabularnewline
$\rho^{\mathfrak{m}}$ & $\rho^{\mathfrak{m}}$ & $-\rho^{\mathfrak{m}}$ & $\rho^{\mathfrak{m}}$\tabularnewline
\hline 
\end{tabular}
\par\end{centering}
\caption{Quaternionic physical quantities and their CPT transformations}
\end{table}
Now, we may apply the CPT transformation relations on generalized
Dirac-Maxwell equations for dyonic fluid of cold plasma as \cite{key-54},

\begin{alignat}{1}
\mathcal{C}PT\left(\boldsymbol{\nabla}\cdotp\boldsymbol{E}\right)T^{-1}P^{-1}\mathcal{C}^{-1}\,\,= & \,\,\mathcal{C}PT\left(\frac{\rho^{e}}{\epsilon}\right)T^{-1}P^{-1}\mathcal{C}^{-1}\,,\nonumber \\
\mathcal{C}PT\left(\boldsymbol{\nabla}\cdotp\boldsymbol{B}\right)T^{-1}P^{-1}\mathcal{C}^{-1}\,\,= & \,\,\mathcal{C}PT\left(\mu\rho^{\mathfrak{m}}\right)T^{-1}P^{-1}\mathcal{C}^{-1}\,,\nonumber \\
\mathcal{C}PT\left(\boldsymbol{\nabla}\times\boldsymbol{B}\right)T^{-1}P^{-1}\mathcal{C}^{-1}\,\,= & \,\,\mathcal{C}PT\left(\frac{1}{a_{0}^{2}}\frac{\partial\boldsymbol{E}}{\partial t}+\mu\boldsymbol{J}^{e}\right)T^{-1}P^{-1}\mathcal{C}^{-1}+\mathcal{C}PT\left(\mu\boldsymbol{J}^{e}\right)T^{-1}P^{-1}\mathcal{C}^{-1}\,,\nonumber \\
\mathcal{C}PT\left(\boldsymbol{\nabla}\times\boldsymbol{E}\right)T^{-1}P^{-1}\mathcal{C}^{-1}\,\,= & \,\,\mathcal{C}PT\left(-\frac{\partial\boldsymbol{B}}{\partial t}\right)T^{-1}P^{-1}\mathcal{C}^{-1}+\mathcal{C}PT\left(-\frac{1}{\epsilon}\boldsymbol{J}^{\mathfrak{m}}\right)T^{-1}P^{-1}\mathcal{C}^{-1}\,.\label{eq:130}
\end{alignat}
Therefore, it may conclude that the generalized Dirac-Maxwell equations
for dyonic cold plasma are invariant under CPT transformations.

\section{Quaternionic hydro-electromagnetic wave propagation}

To establish the dual hydrodynamics wave equations for dyonic cold
plasma, we can start with the following quaternionic relation,
\begin{alignat}{1}
\mathbb{D}\circ(\mathbb{\bar{D}}\circ\boldsymbol{\Psi})\,\,= & \,-\mathbb{D}\circ\mathbb{S}\,,\label{eq:131}
\end{alignat}
where the left hand part of equation (\ref{eq:131}) can be written
as
\begin{alignat}{1}
\mathbb{D}\circ(\mathbb{\bar{D}}\circ\boldsymbol{\Psi})\,\,=\,\,\, & \,e_{1}\left\{ \left(\frac{\partial^{2}B_{x}}{\partial x^{2}}-\frac{1}{a_{0}^{2}}\frac{\partial^{2}B_{x}}{\partial t^{2}}\right)+\frac{i}{a_{0}}\left(\frac{\partial^{2}E_{x}}{\partial x^{2}}-\frac{1}{a_{0}^{2}}\frac{\partial^{2}E_{x}}{\partial t^{2}}\right)\right\} \nonumber \\
+ & \,\,e_{2}\left\{ \left(\frac{\partial^{2}B_{y}}{\partial y^{2}}-\frac{1}{a_{0}^{2}}\frac{\partial^{2}B_{y}}{\partial t^{2}}\right)+\frac{i}{a_{0}}\left(\frac{\partial^{2}E_{y}}{\partial y^{2}}-\frac{1}{a_{0}^{2}}\frac{\partial^{2}E_{y}}{\partial t^{2}}\right)\right\} \nonumber \\
+ & \,\,e_{3}\left\{ \left(\frac{\partial^{2}B_{z}}{\partial z^{2}}-\frac{1}{a_{0}^{2}}\frac{\partial^{2}B_{z}}{\partial t^{2}}\right)+\frac{i}{a_{0}}\left(\frac{\partial^{2}E_{z}}{\partial z^{2}}-\frac{1}{a_{0}^{2}}\frac{\partial^{2}E_{z}}{\partial t^{2}}\right)\right\} \,\,.\label{eq:132}
\end{alignat}
Accordingly, the right hand part of equation (\ref{eq:131}) can be
expressed as
\begin{alignat}{1}
\mathbb{D}\circ\mathbb{S}\,\,= & \,\,e_{\mathrm{1}}\left\{ \mu\left(\frac{\partial J_{z}^{e}}{\partial y}-\frac{\partial J_{y}^{e}}{\partial z}-\frac{1}{a_{0}^{2}\mu\epsilon}\frac{\partial J_{x}^{\mathfrak{m}}}{\partial t}-\frac{\partial\rho^{\mathfrak{m}}}{\partial x}\right)-\frac{i}{a_{0}\epsilon}\left(\frac{\partial J_{z}^{\mathfrak{m}}}{\partial y}-\frac{\partial J_{y}^{\mathfrak{m}}}{\partial z}+\mu\epsilon\frac{\partial J_{x}^{e}}{\partial t}+\frac{\partial\rho^{e}}{\partial x}\right)\right\} \nonumber \\
+ & \,\,e_{\mathrm{2}}\left\{ \mu\left(\frac{\partial J_{x}^{e}}{\partial z}-\frac{\partial J_{z}^{e}}{\partial x}-\frac{1}{a_{0}^{2}\mu\epsilon}\frac{\partial J_{y}^{\mathfrak{m}}}{\partial t}-\frac{\partial\rho^{\mathfrak{m}}}{\partial y}\right)-\frac{i}{a_{0}\epsilon}\left(\frac{\partial J_{x}^{\mathfrak{m}}}{\partial z}-\frac{\partial J_{z}^{\mathfrak{m}}}{\partial x}+\mu\epsilon\frac{\partial J_{y}^{e}}{\partial t}+\frac{\partial\rho^{e}}{\partial y}\right)\right\} \nonumber \\
+ & \,\,e_{\mathrm{3}}\left\{ \mu\left(\frac{\partial J_{y}^{e}}{\partial x}-\frac{\partial J_{x}^{e}}{\partial y}-\frac{1}{a_{0}^{2}\mu\epsilon}\frac{\partial J_{z}^{\mathfrak{m}}}{\partial t}-\frac{\partial\rho^{\mathfrak{m}}}{\partial z}\right)-\frac{i}{a_{0}\epsilon}\left(\frac{\partial J_{y}^{\mathfrak{m}}}{\partial x}-\frac{\partial J_{x}^{\mathfrak{m}}}{\partial y}+\mu\epsilon\frac{\partial J_{z}^{e}}{\partial t}+\frac{\partial\rho^{e}}{\partial z}\right)\right\} \nonumber \\
- & \,\,e_{0}\left\{ \mu\left(\frac{\partial J_{x}^{e}}{\partial x}+\frac{\partial J_{y}^{e}}{\partial y}+\frac{\partial J_{z}^{e}}{\partial z}+\frac{1}{a_{0}^{2}\mu\epsilon}\frac{\partial\rho^{e}}{\partial t}\right)-\frac{i}{a_{0}\epsilon}\left(\frac{\partial J_{x}^{\mathfrak{m}}}{\partial x}+\frac{\partial J_{y}^{\mathfrak{m}}}{\partial y}+\frac{\partial J_{z}^{\mathfrak{m}}}{\partial z}+\mu\epsilon\frac{\partial\rho^{\mathfrak{m}}}{\partial t}\right)\right\} \,\,.\label{eq:133}
\end{alignat}
Now, equating the real and imaginary parts of quaternionic basis vectors
in equation (\ref{eq:131}), and obtained the following relations
\begin{alignat}{1}
\boldsymbol{\nabla}\cdot\boldsymbol{J}^{e}+\frac{\partial\rho^{e}}{\partial t} & \,\,=\,\,0\,,\label{eq:134}\\
\boldsymbol{\nabla}\cdot\boldsymbol{J}^{\mathfrak{m}}+\frac{1}{a_{0}^{2}}\frac{\partial\rho^{\mathfrak{m}}}{\partial t} & \,\,=\,\,0\,,\label{eq:135}
\end{alignat}

\begin{alignat}{1}
\boldsymbol{\nabla}^{2}\boldsymbol{B}-\frac{1}{a_{0}^{2}}\frac{\partial^{2}\boldsymbol{B}}{\partial t^{2}}-\mu\left(\boldsymbol{\nabla}\rho^{\mathfrak{m}}\right)-\frac{1}{a_{0}^{2}\epsilon}\frac{\partial\boldsymbol{J}^{\mathfrak{m}}}{\partial t}+\mu\left(\boldsymbol{\nabla}\times\boldsymbol{J}^{e}\right)\,\, & =\,\,0\,,\label{eq:136}\\
\boldsymbol{\nabla}^{2}\boldsymbol{E}-\frac{1}{a_{0}^{2}}\frac{\partial^{2}\boldsymbol{E}}{\partial t^{2}}-\frac{1}{\epsilon}\left(\boldsymbol{\nabla}\rho^{e}\right)-\mu\frac{\partial\boldsymbol{J}^{e}}{\partial t}-\frac{1}{\epsilon}\left(\boldsymbol{\nabla}\times\boldsymbol{J}^{\mathfrak{m}}\right)\,\, & =\,\,0\,.\label{eq:137}
\end{alignat}
Equations (\ref{eq:134}) and (\ref{eq:135}) are defined the well-known
dual continuity equations while equations (\ref{eq:136}) and (\ref{eq:137})
are represented the generalized hydro-magnetic and hydro-electric
wave equations for dyonic cold plasma in presence of electrons and
magnetic monopoles. The beauty of equation (\ref{eq:136}) is that,
it is an analogous to Alfven wave propagation \cite{key-55,key-56}
associated with magnetic monopoles, and the same way equation (\ref{eq:137})
describes the counterpart of Alfven wave propagation associated with
the electrons. Thus, the unified hydro-electromagnetic wave equations
for dyonic fluid of cold plasma can also be expressed as 

\begin{alignat}{1}
\boldsymbol{\nabla}^{2}\boldsymbol{\Psi}-\frac{1}{a_{0}^{2}}\frac{\partial^{2}\boldsymbol{\Psi}}{\partial t^{2}}-i\,\left(\boldsymbol{\nabla}\text{\ensuremath{\wp}}\right)-\frac{i}{a_{0}}\frac{\partial\boldsymbol{S}}{\partial t}+\left(\boldsymbol{\nabla}\times\boldsymbol{S}\right)\,\, & =\,\,0\,.\label{eq:138}
\end{alignat}
Interestingly, the generalized wave equation (\ref{eq:138}) is invariant
under the duality, Lorentz and CPT transformations.

\section{Conclusion}

The dyons are high energetic soliton particles existed in the cold
plasma. The cold plasma model is the simplest model where we assume
negligible plasma temperature, and the corresponding distribution
function shows the Dirac delta function centered at the macroscopic
flow of linearised velocity. Dyonic cold plasma model can be used
in the study of small amplitude electromagnetic waves propagating
in the conducting plasma. In this study, we have applied the four-dimensional
space-time algebra (quaternionic algebra) to elaborate the dynamics
of dyonic fluid in cold plasma field. In section-2, we have explained
in detail the properties of quaternionic algebra. However, the quaternion
is an important and appropriate fundamental mathematical tool to understand
the four-dimension space-time world. In section-3 \& 4, the fundamental
equations for MHD field and their cold plasma approximation have been
defined. The interesting part we have mentioned here that the dual
MHD equations for massive dyons consisted with electrons and magnetic
monopoles. The generalized equations involving the mass and charge
densities are expressed in terms of one-fluid theory of dyonic cold
plasma. Accordingly, we have discussed the dual current densities
given by equation (\ref{eq:56}). The mass conservation law, dual-charge
conservation law, Lorentz force equation and Ohm's law for dyonic
cold plasma have been defined. In section-5, we have described the
quaternionic formulation for moving massive dyonic fluid of incompressible
cold plasma. The advantage of the quaternionic formulation is that,
it is better to explain two four-velocities, hydro-electric (Lamb
vector) and hydro-magnetic (vorticity) fields and the dual Lorenz
gauge conditions for dyonic cold plasma. It has been emphasized that
the dual hydrodynamics field of dyons (i.e., hydro-electric and hydro-magnetic
fields) deal with both electro-hydrodynamic and magnetic-hydrodynamics.
In present study, the existence of magnetic monopoles has been visualized
to MHD field. It has been shown that the two current sources are also
associated with the quaternionic hydro-electric and hydro-magnetic
fields of dyonic plasma fluid. We have established the eight primary
equations of dual MHD field in presence of dyonic fluid. Interestingly,
the unified macroscopic Dirac-Maxwell equations (\ref{eq:104}), (\ref{eq:105})
have been obtained in the case of dyonic dual MHD. It has been noticed
that like electrodynamics, the Dirac-Maxwell fluid equations are mandatory
to describe the dynamics of MHD plasma. The beauty of cold plasma
field equations is that, these equations are well invariant under
the duality, Lorentz and CPT transformations. In section-6, we have
obtained the quaternionic dual continuity equations for incompressible
dyonic fluid. The generalized hydro-electric and hydro-magnetic wave
equations have been established for dyonic cold plasma in presence
of electrons and magnetic monopoles. It has been emphasized that,
the obtained Alfven wave like equation associated with magnetic monopoles,
while the counterpart of Alfven wave equation plays as electric-plasma
waves in presence of electrons.


\begin{thebibliography}{10}
\bibitem{key-1} M. P. Bachynski, \textbf{``Plasma Physics: An Elementary
Review''}, Proc. IEEE. 49 (1961), 1751.

\bibitem{key-2} L. Tonks, and I. Langmuir, \textbf{``A General Theory
of the Plasma of an Arc'', }Am. Phy. Soc. XXIV (1929), 176.

\bibitem{key-3} H. Alfven, \textbf{``Existence of electromagnetic-hydrodynamic
waves'',} Nature 150 (1942), 405.

\bibitem{key-4}J. D. Jackson, \textbf{``Classical Electrodynamics'',}
3$^{\text{rd}}$ Ed., John Wiley \& Sons, NY(1998), 467.

\bibitem{key-5}J. M. Dawson, \textbf{``Nonlinear Electron Oscillations
in a cold plasma''}, Phys. Rev. 113 (1959), 383.

\bibitem{key-6} N. Meyer- Vernet, \textbf{``Electromagnetic waves
in plasma containing both electric charges and magnetic monopoles''},
Am. J. Phys. \textbf{50} (1981), 846.

\bibitem{key-7} P. A. M. Dirac, ``\textbf{Quantized Singularities
in the Electromagnetic Field}'', Proc. Roy. Soc. London, \textbf{A133}
(1931), 60.

\bibitem{key-8} P. A. M. Dirac, \textbf{``The theory of magnetic
poles'',} Phys. Rev., \textbf{74} (1948), 817.

\bibitem{key-9}J. Schwinger, ``\textbf{A Magnetic Model of Matter}'',
Science, \textbf{165 }(1969), 757.

\bibitem{key-10}J. Schwinger, \textbf{``Sources and Magnetic charge'',}
Phys. Rev., \textbf{173 }(1968), 1536.

\bibitem{key-11}D. Zwanziger, ``\textbf{Dirac Magnetic Poles Forbidden
in S-Matrix Theory}'', Phys. Rev., \textbf{137B} (1965), 647.

\bibitem{key-12}A. Peres, ``\textbf{Rotational invariance of magnetic
monopoles}'', Phys. Rev., \textbf{167} (1968), 1449.

\bibitem{key-13}A. Peres, ``\textbf{Singular String of Magnetic
Monopoles}'', Phys. Rev. Lett., \textbf{18} (1967), 50.

\bibitem{key-14}L. E. Dickson, \textbf{``On Quaternions and their
Generalization and the history of the Eight Square Theorem''}. Ann.
Math., \textbf{20} (1919), 153.

\bibitem{key-15}W. R. Hamilton, \textbf{``Elements of Quaternions''},
Vol. I \& II, Chelsea Publishing, New York (1969), 1185.

\bibitem{key-16}A. Cayley, ``\textbf{On certain results relating
to quaternions}'', Phil. Mag., \textbf{26 }(1845), 210.

\bibitem{key-17}H. T. Flint, \textbf{``Applications of quaternions
to the theory of relativity}'', Phil. Mag., \textbf{39} (1920), 439.

\bibitem{key-18}B. S. Rajput, \textbf{``Unification of generalized
electromagnetic and gravitational fields'',} J. Math. Phys. \textbf{25}
(1984), 351. 

\bibitem{key-19}O. P. S. Negi and B. S. Rajput, \textbf{``Quaternionic
formulation for electromagnetic-field equations''}, Lett. Nuovo Cimento,\textbf{
37} (1983), 325.

\bibitem{key-20}K. Imaeda, \textbf{``Quaternionic formulation of
tachyons, superluminal transformations and a complex space-time''},
Lett. Nuovo Cimento,\textbf{ 50} (1979), 271.

\bibitem{key-21}S. L. Adler, \textbf{``Quaternionic Quantum Mechanics
and Quantum Fields''}, Oxford University Press, New York, (1995).

\bibitem{key-22}A. I. Arbab, \textquotedblleft \textbf{A Quaternionic
Quantum Mechanics}\textquotedblright , Appl. Phys. Res., \textbf{3}
(2011), 160.

\bibitem{key-23}S. Demir, M. Tanisli, \textbf{\textquotedblleft A
compact biquaternionic formulation of massive field equations in gravi
electromagnetism\textquotedblright ,} Eur. Phys. J. Plus, \textbf{126}
(2011), 115.

\bibitem{key-24}S. Demir, M. Tanisli and M. E. Kansu, \textbf{\textquotedblleft Octonic
massless field equations\textquotedblright ,} Int. J. Mod. Phys. A,
\textbf{30} (2015), 1550084.

\bibitem{key-25}\textcolor{black}{B. C. Chanyal, P. S. Bisht and
O. P. S. Negi,}\textit{ }\textbf{``Generalized Octonion Electrodynamics'',}
Int. J. Theor. Phys., \textbf{49} (2010), 1333.

\bibitem{key-26}\textcolor{black}{B. C. Chanyal, P. S. Bisht and
O. P. S. Negi, }\textbf{\textcolor{black}{``Octonion and conservation
laws for dyons'',}}\textcolor{black}{{} Int. J. Mod. Phys. A, }\textbf{\textcolor{black}{28}}\textcolor{black}{{}
(2013), 1350125.}

\bibitem{key-27}B. C. Chanyal, \textbf{``Split octonion reformulation
of generalized linear gravitational field equations}'', J. Math.
Phys., \textbf{56} (2015), 051702.

\bibitem{key-28}S. V. Mironov, V. L. Mironov, \textbf{``Sedeonic
equations of massive fields''}, Int. J. Theor. Phys., \textbf{54}
(2015), 153.

\bibitem{key-29} Z. Weng, \textbf{``Some properties of dark matter
in the complex octonion space''}, \textcolor{black}{Int. J. Mod.
Phys. A,} \textbf{30} (2015), 1550212.

\bibitem{key-30}B. C. Chanyal, \textbf{``A relativistic quantum
theory of dyons wave propagation'', }Canadian J. Phys., \textbf{95}
(2017), 1200.

\bibitem{key-31}B. C. Chanyal, \textbf{``A new development in quantum
field equations of dyons'', }Canadian J. Phys., (2018) (Online Published)
DOI: 10.1139/cjp-2017-0996.

\bibitem{key-32}R. J. Thompson and T. M. Moeller, \textbf{``A Maxwell\textquoteright s
formulation for the equations of a Plasma''}, Phys. Plasmas, \textbf{19}
(2012), 010702.

\bibitem{key-33} S. Demir, M. Tan\i \c{s}l\i , N. \c{S}ahin and M.E.
Kansu, ``\textbf{Biquaternionic reformulation of multifluid plasma
equations}'', Chinese J. Phys., \textbf{55} (2017), 1329.

\bibitem{key-34}S. Demir and E. Zeren, ``\textbf{Multifluid plasma
equations in terms of hyperbolic octonions}'', Int. J. Geom. Meth.
Mod. Phys., \textbf{15} (2017), 1850053.

\bibitem{key-35} D. R. Nickolson, \textbf{``An Introduction to Plasma
Theory''}, John Wiley \& Sons,\textit{ }New York (1983), 171.

\bibitem{key-36} Yu. L. Klimontovich, \textbf{``The statistical
theory of non-equilibrium processes in a plasma''}, M.I.T. Press,
Cambridge, Mass, (1967).

\bibitem{key-37} A. A. Vlasov, ``\textbf{On the Kinetic theory of
an Assembly of particles with collective interaction}'', J. Phys.
(U.S.S.R.) \textbf{9} (1945), 25.

\bibitem{key-38} P. A. Davidson, \textbf{``An Introduction to Magneto-hydrodynamics''},
Cambridge University Press\textit{, }New York (2001), 55.

\bibitem{key-39} P. M. Bellen, \textbf{``Fundamentals of Plasma
Physics''},\textit{ }Cambridge University Press, New York\textit{
}(2006).

\bibitem{key-40} R. Fitzpatrick, \textbf{``Plasma Physics: An Introduction''},
CRC Press, Taylor \& Francis Group, New York (2015).

\bibitem{key-41} M. Goossens, \textbf{``An Introduction to Plasma
Astrophysics and Magneto-hydrodynamics''}, Spring. Sci. \& Business
Media, (2012).

\bibitem{key-42}O. Coceal, W. A. Sabra and S. Thomas, ``\textbf{Duality-invariant
magnetohydrodynamics and dyons}'', EPL, \textbf{35} (1996), 277.

\bibitem{key-43} R. J. Thompson and T. M. Moeller, \textbf{``Classical
field isomorphisms in two-fluid plasmas'',} Phys. Plasma, \textbf{19}
(2012), 082116.

\bibitem{key-44} A. I. Arbab, \textbf{``The analogy between electromagnetism
and hydrodynamics''}, Phys. Essays, \textbf{24} (2011), 2.

\bibitem{key-45} H. Lamb, \textbf{``Hydrodynamics''}, Cambridge
University Press, New York (1932), 134.

\bibitem{key-46} C. W. Hamman, J. C. Klewicki and R. M. Kirby, \textbf{``On
the Lamb vector divergence in Navier-Stokes flows''}, J. Fluid Mech.,
\textbf{610} (2008), 261.

\bibitem{key-47} C. Truesdell, \textbf{``The kinematics of vorticity''}
(Vol.954), Bloomington: Indiana University Press (1954).

\bibitem{key-48}N. Anderson and A. M. Arthurs, \textbf{``Duality
transformation and invariants of the electromagnetic field}'', Int.
J. Electronics, \textbf{69} (1990), 575.

\bibitem{key-49} R. D. Sard, \textbf{``Relativistic mechanics: Special
relativity and Classical particle dynamics''}, W. A. Benjamin, New
York (1970).

\bibitem{key-50} C. A. Brau, \textbf{``Modern Problems to Classical
Electrodynamics''},\textit{ }Oxford University Press, New York (2003).

\bibitem{key-51} B. C. Chanyal, \textbf{``Octonion massive electrodynamics''},
Gen. Relativ. Gravit. \textbf{46} (2014), 16461. 

\bibitem{key-52} J. W. Norbury, \textbf{``The invariance of classical
electromagnetism under charge conjugation, parity and time reversal
(CPT) transformations''}, Eur. J. Phys., \textbf{11} (1990), 99.

\bibitem{key-53}D. Malament, \textbf{``On the time reversal invariance
of classical electromagnetic theory'',} Stud. Hist. Philos. Mod.
Phys. \textbf{35B} (2004), 295.

\bibitem{key-54}P. S. Bisht, T. Li, Pushpa, and O. P. S. Negi, \textbf{``Discrete
symmetries and generalized fields of dyons}'', Inter. J. Theor. Phys,
\textbf{49} (2010), 1370.

\bibitem{key-55} A. Hasegawa, and C. Uberoi, \textbf{``The Alfven
Waves''}, DOE Critical Rev. Series, Tech. Information center, USA\textit{
}(1982).

\bibitem{key-56}H. C. Spruit, \textbf{``Essential magnetohydrodynamics
for astrophysics''}, e-print\textit{ }arXiv:1301.5572 {[}astro-ph.IM{]}\textit{
}(2016).
\end{thebibliography}
\end{document}